\documentclass[12pt]{iopart}

\usepackage{graphicx}
\usepackage{graphics,epsf}
\usepackage{colordvi}
\begin{document}

\title[On the selective multiphoton ionization of sodium by
femtosecond laser pulses]{On the selective multiphoton ionization
of sodium by femtosecond laser pulses: A partial-wave analysis}

\author{A. Bunjac, D. B. Popovi\'c and N. S. Simonovi\'c}
\address{Institute of Physics, University of Belgrade, P.O. Box 57,
11001 Belgrade, Serbia}


\begin{abstract}
Multiphoton ionization of sodium by femtosecond laser pulses of
800\,nm wavelength in the range of laser peak intensities entering
over-the-barrier ionization domain is studied. Photoelectron
momentum distributions and the energy spectra are determined
numerically by solving the time dependent Schr\"odinger equation
for three values of the laser intensity from this domain. The
calculated spectra agree well with the spectra obtained
experimentally by Hart {\em et al}~({\em Phys.~Rev.}~A 2016
\textbf{93} 063426). A partial wave analysis of the spectral peaks
related to Freeman resonances has shown that each peak is a
superposition of the contributions of photoelectrons produced by
the resonantly enhanced multiphoton ionization via different
intermediate states. It is demonstrated that at specific laser
intensities the selective ionization, which occurs predominantly
through a single intermediate state, is possible.
\end{abstract}
%
%
%
%
%



\section{Introduction}
\label{intro}

Strong-field ionization of the alkali-metal atoms has been studied
intensively over the past decade and earlier, both experimentally
and theoretically including {\em ab initio} numerical calculations
\cite{wollenhaupt,krug,schuricke, JJ, morishita, schuricke2,
hart2016, pccp, wessels}. A specific feature of this group of
atoms -- a low ionization potential, which ranges from $I_p
\approx 3.89$\,eV (for cesium) to 5.39\,eV (for lithium), causes
that a considerably smaller number of photons of a given energy
$\hbar\omega$ is required for their photoionization than for the
ionization of other atoms. For example, with the laser wavelength
of around 800 nm ($\hbar \omega \approx 1.55$\,eV) it takes four
photons to ionize an alkali-metal atom, unlike the case of
frequently used noble gases where this number is of the order of
ten. Since for a dipole transition requiring $N$ photons the
lowest order perturbation theory predicts that the photon
absorption rate is $W \sim I^N$ if $I \ll I_a$, where $I$ is the
laser intensity and $I_a = 3.50945 \times
10^{16}\,\mathrm{W/cm}^2$ is the atomic unit value for intensity,
measurable effects in experiments with multiphoton ionization
(MPI) of alkali can be observed at relatively low laser
intensities, available in table-top laser systems.

The perturbative  treatment, however, is not applicable at higher
intensities which can be achieved today. One indication of the
nonperturbative regime is the so-called above threshold ionization
(ATI) \cite{mittleman,dk2000,JKP} in which the atom absorbs more
photons than the minimum required. Under these conditions the
photoelectron energy spectra (PES, electron yield versus their
excess energy $\epsilon$) were seen to consist of several peaks,
separated by the photon energy $\hbar\omega$, and appearing at
energies $\epsilon = (N_0 + s)\hbar\omega - I_p$, where $N_0$ is
the minimum number of photons needed to exceed the ionization
potential $I_p$ and $s = 0,1,\ldots$ is the number of excess
("above-threshold") photons absorbed by the atom. (For the
alkali-metal atoms and the laser of 800\,nm wavelength one has
$N_0 = 4$.) By increasing the intensity over a certain value, $W$
does not follow further the prediction $I^{N_0+s}$ of the
perturbation theory.

At even larger intensities, the electric component of the laser
field becomes comparable with the atomic potential, opening up
another ionization mechanism -- the tunnel ionization. In this
case the field distorts the atomic potential forming a potential
barrier through which the electron can tunnel. Multiphoton and
tunneling ionization regimes are distinguished by the value of
Keldysh parameter \cite{keldysh} which can be written as $\gamma =
\sqrt{I_p/(2U_p)}$, where $U_p = e^2F^2/(4m_e\omega^2)$ is the
ponderomotive potential of ejected electron with mass $m_e$ and
charge $e$. The value of the electric field $F$ in the expression
for $\gamma$ corresponds to the peak value of laser intensity.
Multiphoton and tunneling regimes are characterized by $\gamma \gg
1$ (high-intensity, long-wavelength limit) and $\gamma \ll 1$
(low-intensity, short-wavelength limit), respectively. The
transition regime at $\gamma \approx 1$ for alkali-metal atoms is
reached at considerably lower intensities than for other atoms,
again due to the small ionization potential $I_p$. The experiments
accessing the strong-field regime with alkali
\cite{schuricke,schuricke2,hart2016,wessels} have revealed that
the commonly used strong-field ionization models in the form of a
pure MPI or tunnel ionization cannot be strictly applied. The
problem, however, goes beyond by using an {\em ab-initio}
numerical method for solving the time-dependent Schr\"odinger
equation (TDSE).

Finally, at a sufficiently high laser intensity, the field
strength overcomes the atomic potential. This can be considered as
the limiting case of tunnel ionization when the barrier is
suppressed below the energy of atomic state. This regime is
usually referred to as over-the-barrier ionization (OBI). Such a
barrier suppression takes place independently of the value of
Keldysh parameter. For neutral atoms the threshold value of field
strength for OBI is estimated as $F_\mathrm{OBI} \approx I_p^2/4$
(in atomic units). $F_\mathrm{OBI}$ values for alkali, determined
more accurately, are given in Ref.~\cite{MS}. The corresponding
laser intensities can be obtained by formula $I = I_a F^2$, where
$F$ is expressed in atomic units and $I_a$ is the above introduced
atomic unit for intensity. For noble gas atoms irradiated by the
laser of wavelength from the visible light domain, OBI was
occurring well into the tunneling regime \cite{mevel}. This is,
however, not a general rule. For atoms with low ionization
potentials, as the alkali-metal atoms are, the OBI threshold,
compared to that for hydrogen or noble gases, is shifted to
significantly lower values of the field strength. For example, the
laser peak intensity that corresponds to the OBI threshold for
sodium is about $3.3 \, \mathrm{TW/cm}^2$ ($F_\mathrm{OBI} =
0.0097$\,a.u. \cite{MS}), whereas the value of Keldish parameter
for the sodium atom interacting with the radiation of this
intensity and 800\,nm wavelength is $\gamma = 3.61$. Thus, the OBI
threshold in this case belongs to the MPI regime. Previous
experiments and theoretical studies have already mentioned this
peculiar situation for sodium and other alkali
\cite{schuricke,schuricke2,hart2016,morishita,JJ,wessels}. In
addition, it is demonstrated that at intensities above the OBI
threshold the atomic target is severely ionized before the laser
peak intensity is reached \cite{morishita}. Thus, the ionization
occurs at the leading edge of the pulse only, that is equivalent
to the ionization by a shorter pulse.

A remarkable feature of the photoelectron spectra obtained using
short (sub-picosecond) laser pulses is the existence of
substructures in ATI peaks known as Freeman resonances. The
mechanism which is responsible for occurrence of these
substructures is the dynamic (or AC) Stark shift
\cite{mittleman,dk2000,dk1999} which brings the atomic energy
levels into resonance with an integer multiple of the photon
energy. Freeman et al.~\cite{freeman,gibons} have shown that when
atomic states during the laser pulse transiently shift into
resonance (this occurs in principle twice during the pulse, once
as the laser pulse "turns-on" and again as the pulse "turns-off"),
the resonantly enhanced multiphoton ionization (REMPI)
\cite{dk2000,grossmann,JKP}) takes place, increasing the
photoelectron yield, and one observes peaks at the corresponding
values of photoelectron energy. Thus, the peaks in the PES can be
related to REMPI occurring via different intermediate states.

The resonant dynamic Stark shift of energy levels corresponding to
sodium excited states $nl$ ($n \le 6$), relative to its ground
state (3s) energy, is recently calculated for the laser
intensities up to $7.9\,\mathrm{TW/cm}^2$ and wavelengths in the
range from 455.6 to 1139\,nm \cite{pccp}. These data are used to
predict the positions of REMPI peaks in the PES of sodium
interacting with an 800\,nm laser pulse. Freeman resonances in the
PES of alkali-metal atoms have been studied in papers
\cite{wollenhaupt,krug,schuricke,JJ,morishita,schuricke2,hart2016,pccp,wessels},
mentioned at the beginning of Introduction, where a number of
significant results have been reported.

The dynamic Stark shift also appears as an important mechanism in
the strong-field quantum control of various atomic and molecular
processes \cite{rabitz,shapiro,sussman,g-vazquez}. Focusing on the
MPI of atoms, a particular challenge would be the selective
ionization of an atom through a single intermediate state which
could produce a high ion yield. By increasing simply the laser
intensity one increases the yield, but also spreads the electron
population over multiple states \cite{gibons} and, in turn,
reduces the selectivity. Krug {\em et al} \cite{krug} demonstrated
that chirped pulses can be an efficient tool in strong-field
quantum control of multiple states of sodium at the MPI. Hart {\em
et al} \cite{hart2016} have shown that improved selectivity and
yield could be achieved by controlling the resonant dynamic Stark
shift via intensity of the laser pulse of an appropriate
wavelength ($\sim 800$\,nm).

In this paper we study the photoionization of sodium by the laser
pulse of 800\,nm wavelength and 57\,fs full width at half maximum
(FWHM) with the peak intensities ranging from $3.5$ to
8.8\,TW/cm$^2$, which belong to OBI domain in the MPI regime.
These values were chosen for comparison with the experiment by
Hart {\em et. al.} \cite{hart2016}. Using the
single-active-electron approximation we calculate the
corresponding photoelectron momentum distribution (PMD) and the
PES by solving numerically the TDSE and perform a similar analysis
as it has been done in Refs.
\cite{wollenhaupt,krug,schuricke,JJ,morishita,schuricke2,hart2016,pccp,wessels}.
In order to make a deeper insight into the ionization process, in
addition, we perform a partial-wave analysis of the calculated
PMD. In the next section we describe the model and in
Sec.~\ref{sec:scheme} consider the excitation scheme and
ionization channels. In Sec.~\ref{sec:results} we analyze the
calculated photoelectron momentum distribution and energy spectra.
A summary and conclusions are given in Sec.~\ref{sec:conc}.

\section{The model}
\label{sec:model}

Singly-excited states and the single ionization of the
alkali-metal atoms are, for most purposes, described in a
satisfactory manner using one-electron models. This follows from
the structure of these atoms, which is that of a single valence
electron moving in an orbital outside a core consisting of closed
shells. In that case the valence electron is weakly bound and can
be considered as moving in an effective core potential
$V_\mathrm{core}(r)$, which at large distances $r$ approaches the
Coulomb potential $-1/r$. One of the simplest models for the
effective core potential, applicable for the alkali-metal atoms,
is the Hellmann pseudopotential \cite{hellmann} which reads (in
atomic units)
\begin{equation}
V_\mathrm{core}(r) = - \frac{1}{r} + \frac{A}{r}\,e^{-ar}.
\label{hellmannECP}
\end{equation}
The parameters $A = 21$ and $a = 2.54920$ \cite{MS} provide the
correct value for the ionization potential of sodium $I_\mathrm{p}
= 5.1391\,\mathrm{eV} = 0.18886$\,a.u. and reproduce approximately
the energies of singly-excited states \cite{sansonetti}
(deviations are less than 1\%). The associated eigenfunctions are
one-electron approximations of these states and have the form
$\psi_{nlm}(\mathbf{r}) = R_{nl}(r) Y_{lm}(\Omega)$. Radial
functions $R_{nl}(r)$ can be determined numerically by solving the
corresponding radial equation.

Here we use this single-active-electron (SAE) approximation to
study the single-electron excitations and ionization of the sodium
atom in a strong laser field. Assuming that the field effects on
the core electrons can be neglected (the so-called frozen-core
approximation \cite{MS}), the Hamiltonian describing the dynamics
of valence (active) electron of the sodium atom in an alternating
field, whose electric component is $F(t) \cos\omega t$, reads (in
atomic units)
\begin{equation}
H = -\frac{1}{2}\nabla^2 + V_\mathrm{core}(r) - F(t) z \cos\omega
t. \label{hamiltonian}
\end{equation}
We consider the linearly polarized laser pulse whose amplitude of
the electric field component (field strength) has the form
\begin{equation}
F(t) = F_\mathrm{peak} \sin^2(\pi t/T_\mathrm{p}),\quad 0 < t <
T_\mathrm{p}, \label{pulse}
\end{equation}
otherwise $F(t) = 0$. Here $\omega$, $F_\mathrm{peak}$ and
$T_\mathrm{p}$ are the frequency of the laser field, the peak
value of $F$ and the pulse duration ($2\times \mathrm{FWHM}$),
respectively. Since the system is axially symmetric, the magnetic
quantum number $m$ of the active electron is a good quantum number
for any field strength. In the sodium ground state (when $F = 0$)
the orbital and magnetic quantum numbers are equal to zero and in
our calculations we set $m = 0$.

Photoabsorption processes were simulated by solving numerically
the TDSE for the active electron wave function
$\psi(\mathbf{r},t)$, assuming that at $t = 0$ the atom is in the
ground state represented by the lowest eigenstate of Hamiltonian
(\ref{hamiltonian}) with $F = 0$. The excitation was studied
preliminary by the method of time dependent coefficients (TDC),
where the wave function $\psi(\mathbf{r},t)$ was expanded in a
finite basis consisting of functions $\psi_{nl0}(\mathbf{r})$. In
this approach the populations of atomic states are the squares of
absolute values of expansion coefficients $c_{nl}(t)$, determined
by solving the corresponding set of equations numerically. This
method, however, cannot be used to calculate photoionization
because the basis of atomic states does not include the continuum
states. An adequate alternative for this purpose is the
wave-packet propagation method on a spatial grid. Here we use the
second-order-difference (SOD) scheme \cite{askar} adapted to
cylindrical coordinates $(\rho,\varphi,z)$, see
Refs.~\cite{pccp,epjd2017}. Due to the axial symmetry of the
system, Hamiltonian (\ref{hamiltonian}) and the electron's wave
function do not depend on the azimuthal angle and the dynamics
reduces to two degrees of freedom ($\rho$ and $z$). The
calculations were performed on a $6144 \times 12288$ grid in the
wave-packet propagation domain $\rho, |z| \le L = 3000$\,a.u. and
the propagation time was about $1.3\,T_p$. The absorbing potential
of form $-i\,0.03\, (r-r_0)^2$ in the area $r
> r_0 = L - 100$\,a.u. was used to suppress the reflection of the
wave packet from the domain boundaries.

\section{Energy scheme and photoionization channels}
\label{sec:scheme}

\begin{figure*}[!]
\begin{center}
\includegraphics[scale=.375]{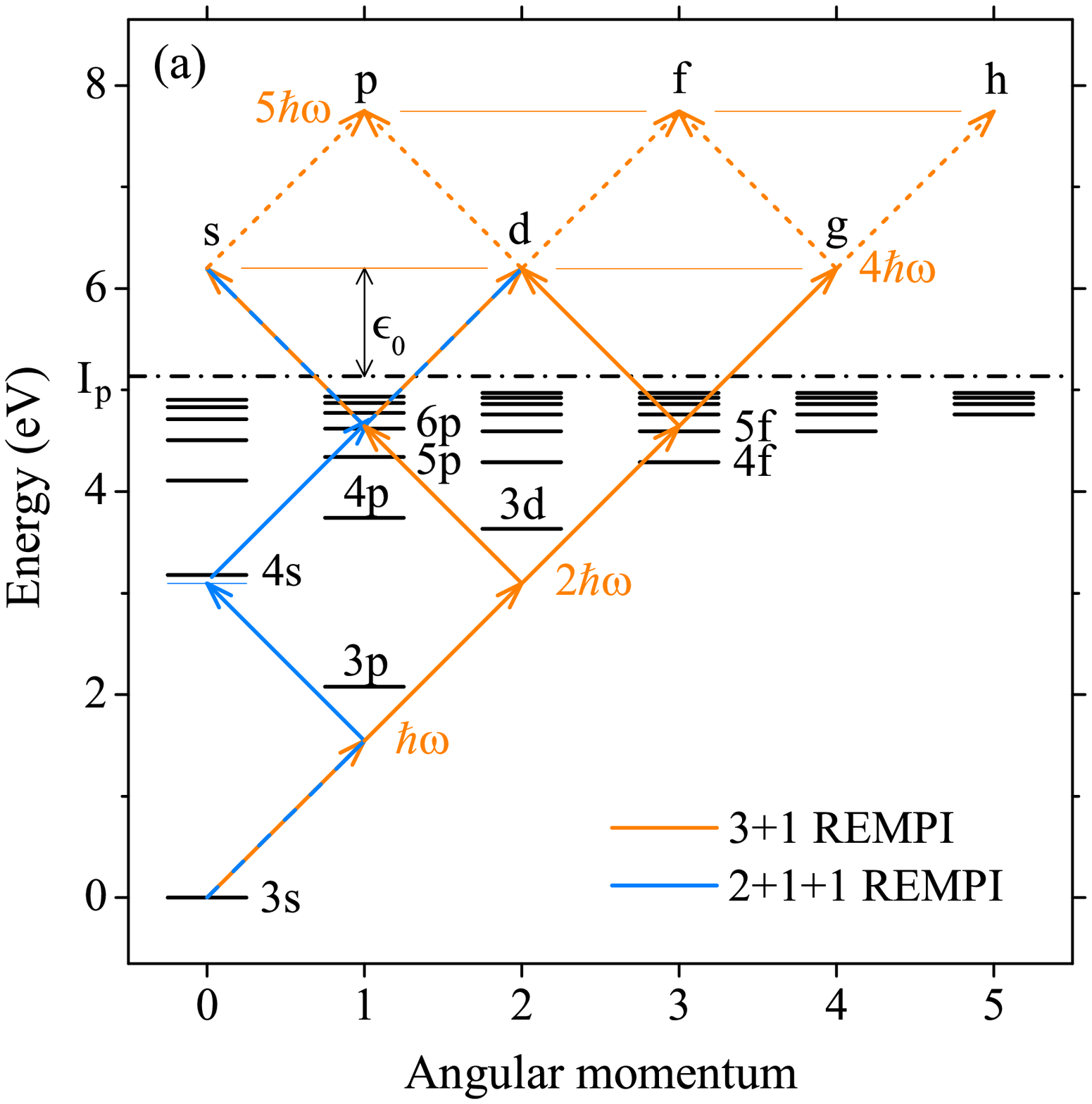}
\includegraphics[scale=.375]{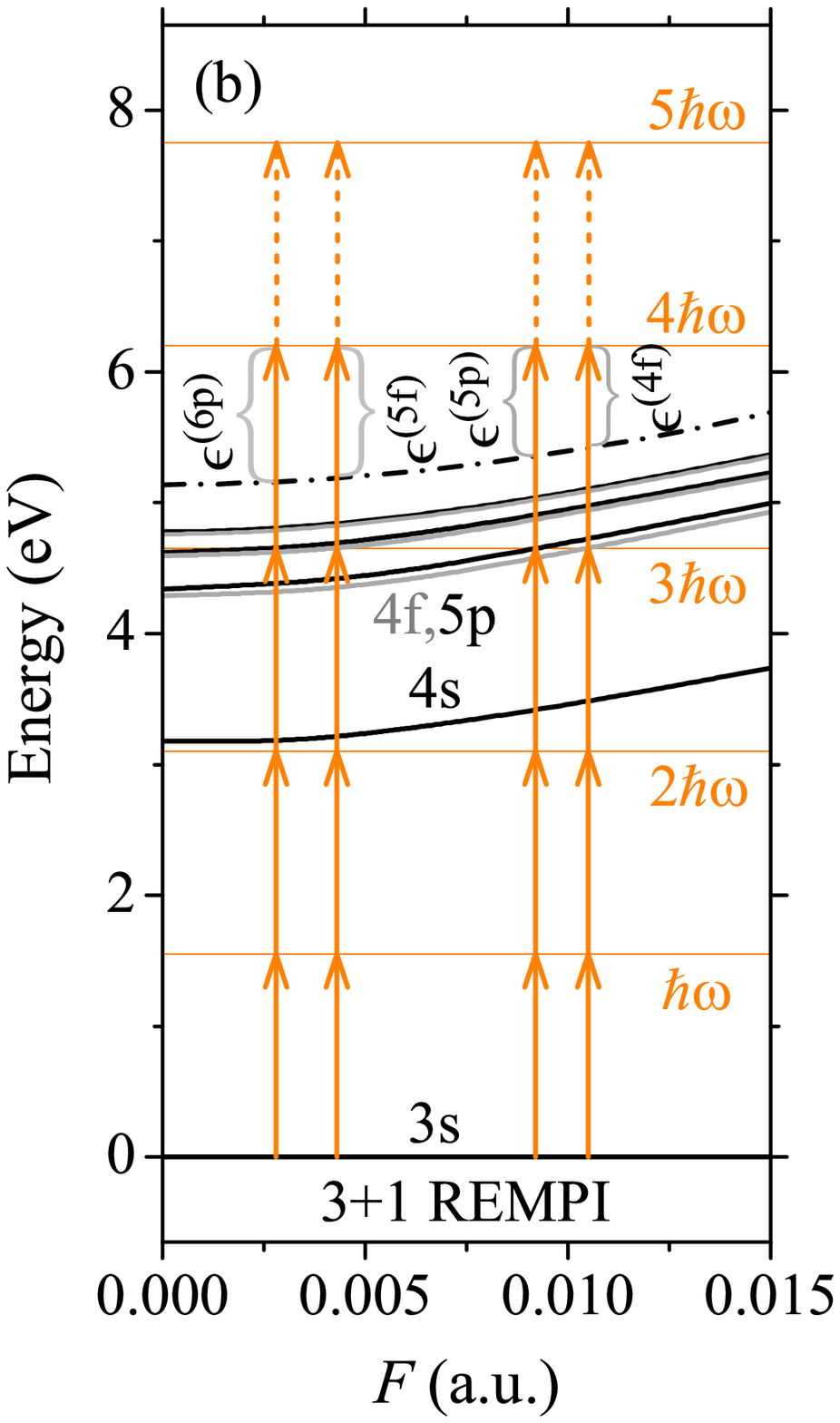}
\includegraphics[scale=.375]{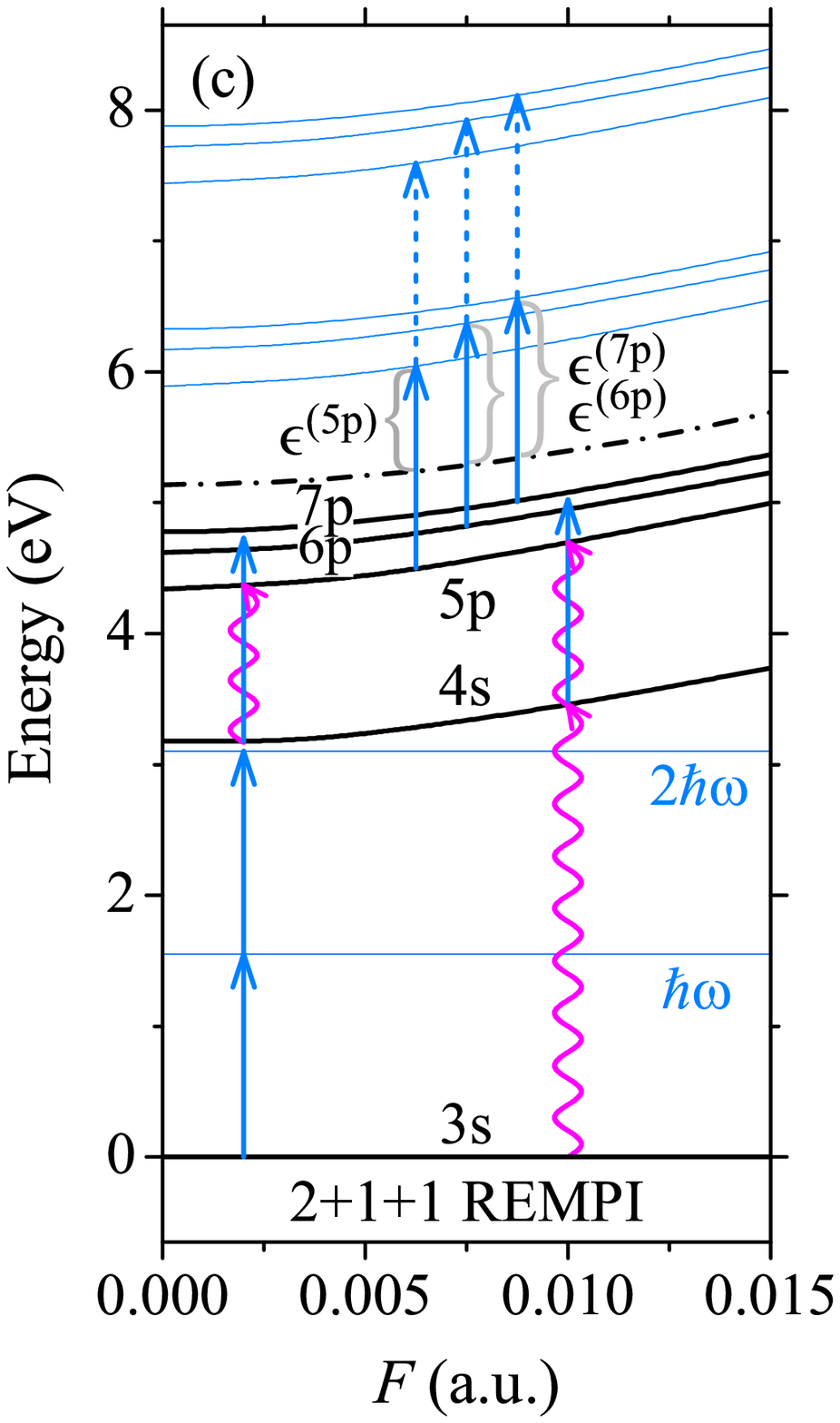}
\end{center}
\caption{(a) The unperturbed energy levels (short black lines)
corresponding to singly excited states of sodium \cite{sansonetti}
relative to its ground state (3s) and possible four-photon and
five-photon absorption pathways (arrows) from the ground state to
continuum for the radiation of 800\,nm wavelength ($\hbar\omega
\approx 1.55\,\mathrm{eV}$). The continuum boundary is drown by
the dash-dot line and $\epsilon_0$ is the excess energy of
photoelectrons produced in the nonresonant four-photon ionization.
(b) Energy levels corresponding to 4s and P states (black curves)
and to F states (gray curves), as functions of the field strength,
and 3+1 REMPI pathways via intermediate states 4f, 5p, 5f and 6p
(orange arrows). $\epsilon^{(nl)}$ are the corresponding
photoelectron excess energies. (c) Energy levels corresponding to
4s and P states as functions of the field strength (black curves)
and 2+1+1 REMPI pathways via near resonant (at weak fields)
$\mathrm{3s} \to \mathrm{4s}$ transition and subsequent excitation
of P states (blue arrows). The wave lines represent nonresonant
transitions related to Rabi oscillations between the corresponding
states.} \label{fig:diagram}
\end{figure*}

The lowest energy levels corresponding to singly-excited states of
sodium and possible multiphoton absorption pathways during the
interaction of the atom with a laser radiation of 800\,nm
wavelength ($\hbar\omega = 0.05695\,\mathrm{a.u.} \approx
1.55\,\mathrm{eV}$) are shown in Fig.~\ref{fig:diagram}(a). At
this wavelength there exist three dominant REMPI channels: (i)
(3+1)-photon ionization via excitation of 5p, 6p and 7p states,
giving rise to photoelectrons with s and d-symmetry; (ii)
(3+1)-photon ionization via excitation of 4f, 5f and 6f states,
producing photoelectrons with d and g-symmetry; (iii)
(2+1+1)-photon ionization via nearly resonant two-photon
transition $3\mathrm{s} \to 4\mathrm{s}$ and subsequent excitation
of P-states, giving rise again to photoelectrons with s and
d-symmetry \cite{krug,hart2016}.

\subsection{Estimation of the photoelectron excess energy. The nonresonant MPI}

Theoretically, if the MPI occurs by absorbing $N$ photons, the
excess energy of ejected electrons in the weak field limit is
$\epsilon^{(0)} = N\!\hbar\omega - I_p$. At stronger fields,
however, the dynamic Stark shift of the ground state ($\delta
E_\mathrm{gr}$), as well as that of the continuum boundary
($\delta E_\mathrm{cb}$), change effectively the ionization
potential $I_p$ to $I_p - \delta E_\mathrm{gr} + \delta
E_\mathrm{cb}$ and the excess energy becomes dependent on the
field strength (see Fig.~\ref{fig:diagram}(b))
\begin{equation}
\epsilon(F) = N\!\hbar\omega - I_p + \delta E_\mathrm{gr}(F) -
\delta E_\mathrm{cb}(F). \label{excess-e-exact1}
\end{equation}
This equation, using quadratic approximation $\delta E \approx
-\alpha(\omega) F^2\!/4$, reads \cite{pccp}
\begin{equation}
\epsilon(F) \approx N\!\hbar\omega - I_p -
\bigg(\!\alpha_\mathrm{gr}^\mathrm{stat} +
\frac{e^2}{m_{\!e\,}\omega^2}\!\bigg)\frac{F^2}{4},
\label{excess-e-approx1}
\end{equation}
where the dynamic polarizability $\alpha(\omega)$ in the ground
state and at the continuum boundary is approximated by its static
value for the sodium ground state
$\alpha_\mathrm{gr}^\mathrm{stat} = 162.7$\,a.u.~\cite{mitroy} and
by its asymptotic value in the high frequency limit
$\alpha_\mathrm{cb}(\omega) \approx -e^2/(m_{\!e\,}\omega^2)$,
respectively. Thus, $\delta E_\mathrm{cb} \approx U_p$, where
$U_p$ is the ponderomotive potential of the active electron,
whereas $\delta E_\mathrm{gr} \approx -0.53\,U_p$.

When the MPI is induced by a laser pulse, the field strength
varies from 0 to $F_\mathrm{peak}$ and back producing
photoelectrons of different energies, but with the maximum yield
at $\varepsilon(F_\mathrm{peak})$ (for a given $N$). This peak in
the PES corresponds to the nonresonant $N$-photon ionization and
its position can be estimated by formula (\ref{excess-e-approx1})
width $F = F_\mathrm{peak}$.

\subsection{3+1 REMPI channels} \label{sec:3+1REMPI}

Formula (\ref{excess-e-approx1}) is also useful for estimating the
positions of REMPI peaks in the PES. Replacing the variable $F$
with the value $F_{nl}$ at which the atomic state $nl$ shifts in
the three-photon resonance with the laser field, this formula for
$N = 4$ estimates the excess energy $\epsilon^{(nl)}$ of
photoelectrons produced in the 3+1 REMPI via this state (see
Fig.~\ref{fig:diagram}(b)). The values $F_{nl}$ for states 4p, 4f,
5p, 5f and 6p in the case of laser field of 800 nm wavelength are
determined in a previous work \cite{pccp}. These values, together
with the corresponding values for $\epsilon^{(nl)}$ estimated by
formula (\ref{excess-e-approx1}) are given in Table~\ref{table1}.

Rewriting the resonance condition $E_{nl}(F_{nl}) -
E_\mathrm{gr}(F_{nl}) = 3\hbar\omega$ in the form $E_{nl}(F_{nl})
= 3\hbar\omega - I_p + \delta E_\mathrm{gr}(F_{nl})$ and inserting
it into Eq.~(\ref{excess-e-exact1}) (with $N = 4$ and $F =
F_{nl}$), one has
\begin{equation}
\epsilon^{(nl)} = E_{nl}(F_{nl}) - \delta E_\mathrm{cb}(F_{nl}) +
\hbar\omega. \label{excess-e-exact2}
\end{equation}
Since the dynamic Stark shift for the high lying levels takes
approximately the same value as that for the continuum boundary
($\delta E_{nl}(F) \approx \delta E_\mathrm{cb}(F) \approx
U_p(F)$), the photoelectron energy at the 3+1 REMPI via considered
state will be
\begin{equation}
\epsilon^{(nl)} \approx E_{nl}^{(0)} + \hbar\omega,
\label{excess-e-approx2}
\end{equation}
where $E_{nl}^{(0)}$ is the energy of the state $nl$ for the
field-free atom. The positions of REMPI maxima in the PES are,
therefore, almost independent on the peak intensity of the laser
pulse, in contrast to the position of the nonresonant four-photon
ionization maximum $\epsilon(F_\mathrm{peak})$. The values for
$\epsilon^{(nl)}$ obtained by Eq.~(\ref{excess-e-approx2}) are
shown in the last column of Table~\ref{table1}. Since usually
$\delta E_{nl} + \delta E_\mathrm{gr} > 0$ (at least for P and F
states, see Fig.~\ref{fig:diagram}(b)), the states which can be
shifted into three-photon resonance are those with $E_{nl}^{(0)}
\le 3\hbar\omega - I_p$ (for the wavelength of 800\,nm these
states are 4f, 5p, 5f, 6p, but not 7p and 6f, which are only near
resonant at small values of $F$). As a consequence the REMPI
maxima are in the spectrum located below the theoretical value for
photoelectron energy in the weak field limit ($\epsilon^{(nl)} \le
\epsilon^{(0)}$).

\begin{table}
\caption{\label{table1} Energies $E_{nl}^{(0)}$ of singly excited
P and F-states ($nl$ from 4p to 7p) of the field free sodium atom
\cite{sansonetti}, field strengths $F_{nl}$ at which these states
shift into the three-photon resonance with the laser field of
800\,nm wavelength ($E_{nl}(F_{nl}) - E_\mathrm{3s}(F_{nl})=
3\hbar\omega$, $\hbar\omega \approx 1.55$\,eV) \cite{pccp} and the
excess energies $\epsilon^{(nl)}$ of photoelectrons produced in
the 3+1 REMPI via these states, obtained by
Eq.~(\ref{excess-e-approx1}). The values for $\epsilon^{(nl)}$
obtained by formula (\ref{excess-e-approx2}), which are also
related to 2+1+1 REMPI via 4s and subsequent excitation of P
intermediate states, are shown in the fifth column.}
  \begin{indented}
    \item[]\begin{tabular}{@{}ccccc} \br
    state ($nl$) & $E_{nl}^{(0)}$\,(eV) & $F_{nl}$\,(a.u.) &
    $\epsilon^{(nl)}$\,(eV) & $E_{nl}^{(0)} \!+\! \hbar\omega$\,(eV) \\
    \mr
    4p & $-1.386$ & 0.0148 & 0.358 & (0.164) \\
    4f & $-0.851$ & 0.0105 & 0.707 & 0.699 \\
    5p & $-0.795$ & 0.0092 & 0.789 & 0.755 \\
    5f & $-0.545$ & 0.0043 & 1.001 & 1.005 \\
    6p & $-0.515$ & 0.0028 & 1.035 & 1.035 \\
    6f & $-0.378$ & -- & -- & 1.172 \\
    7p & $-0.361$ & -- & -- & 1.189 \\
    \br
  \end{tabular}
  \end{indented}
\end{table}

\subsection{The 2+1+1 REMPI channel} \label{sec:2+1+1REMPI}

\begin{figure*}
\begin{center}
\includegraphics[width=.45\textwidth]{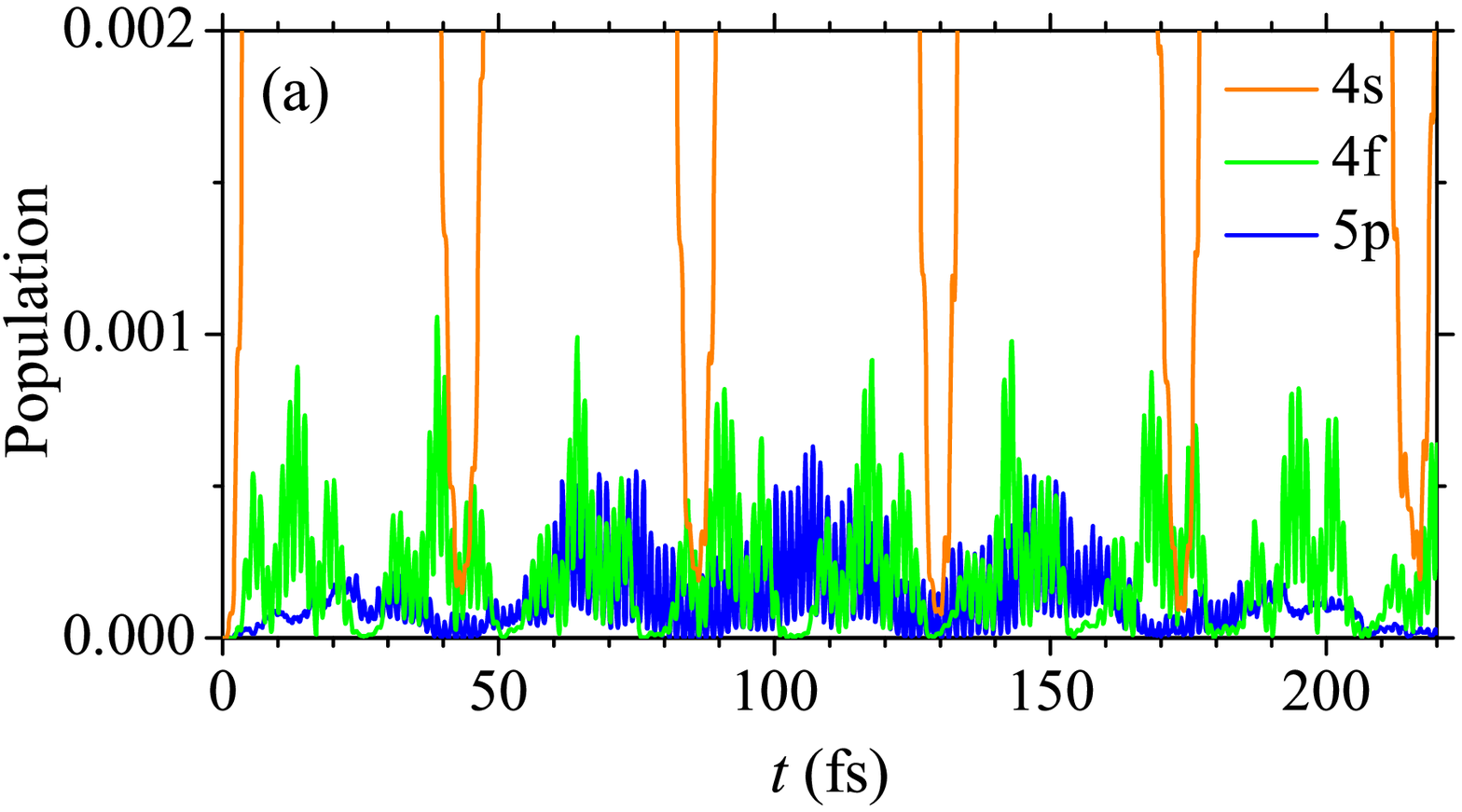}
\includegraphics[width=.45\textwidth]{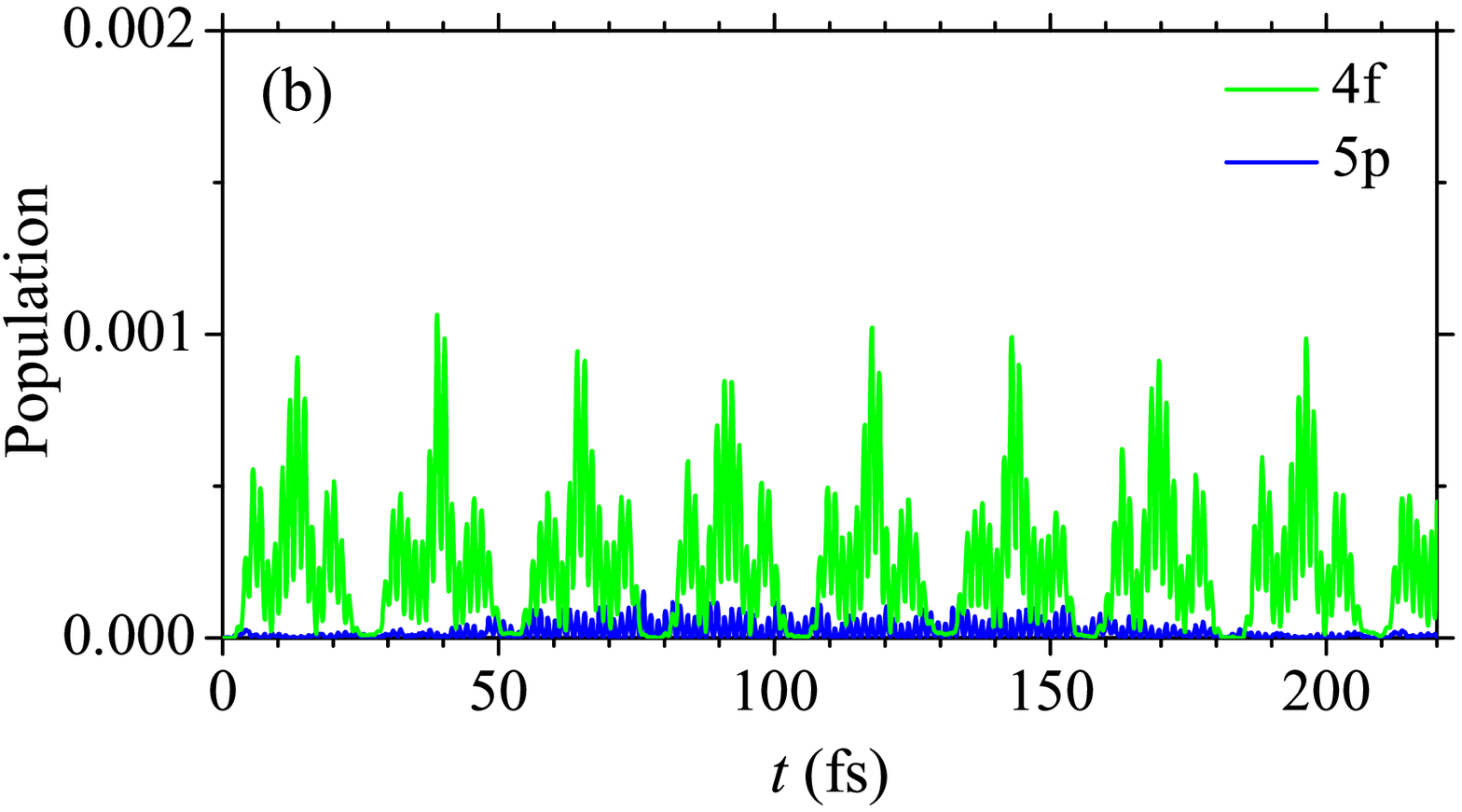}
\\
\includegraphics[width=.45\textwidth]{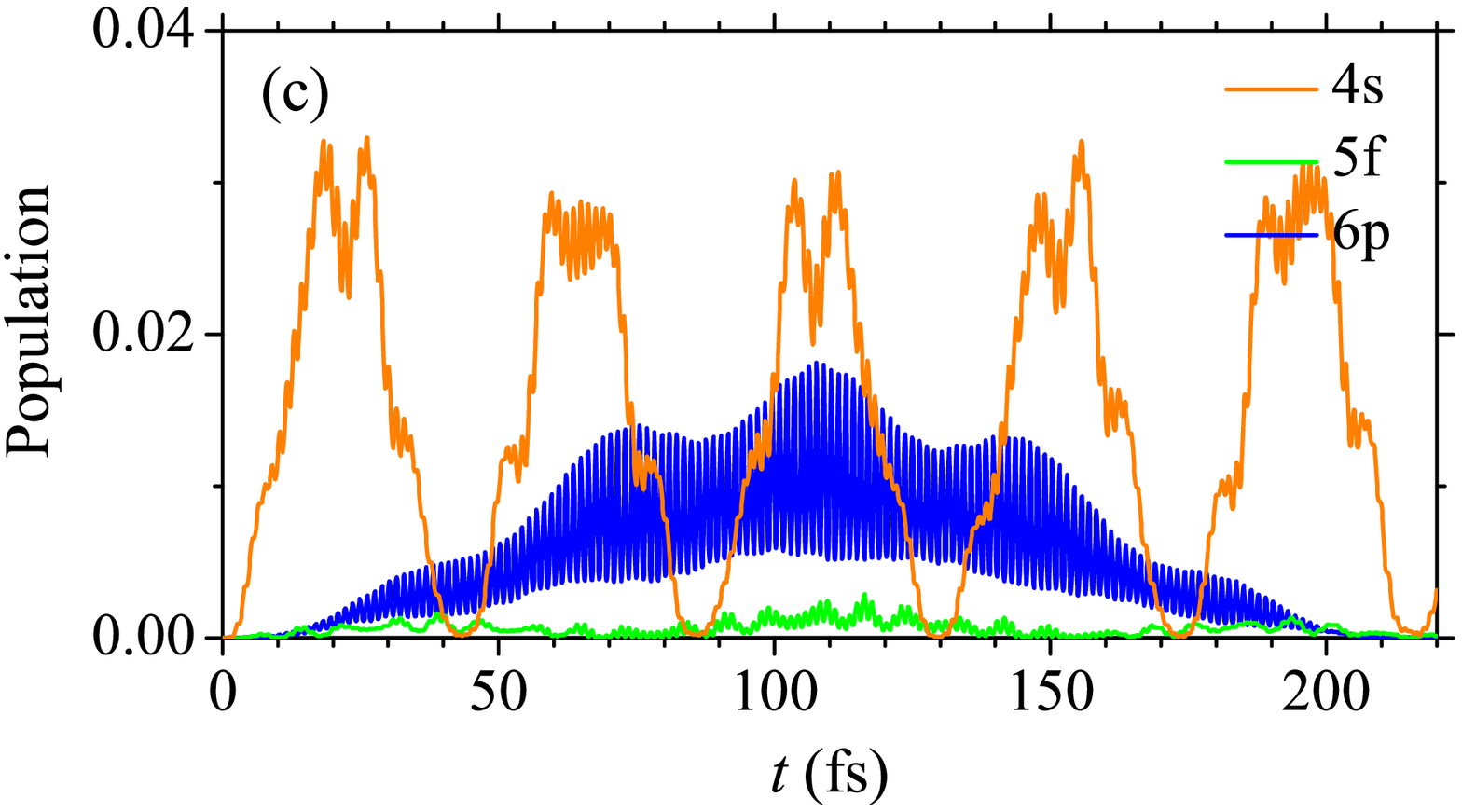}
\includegraphics[width=.45\textwidth]{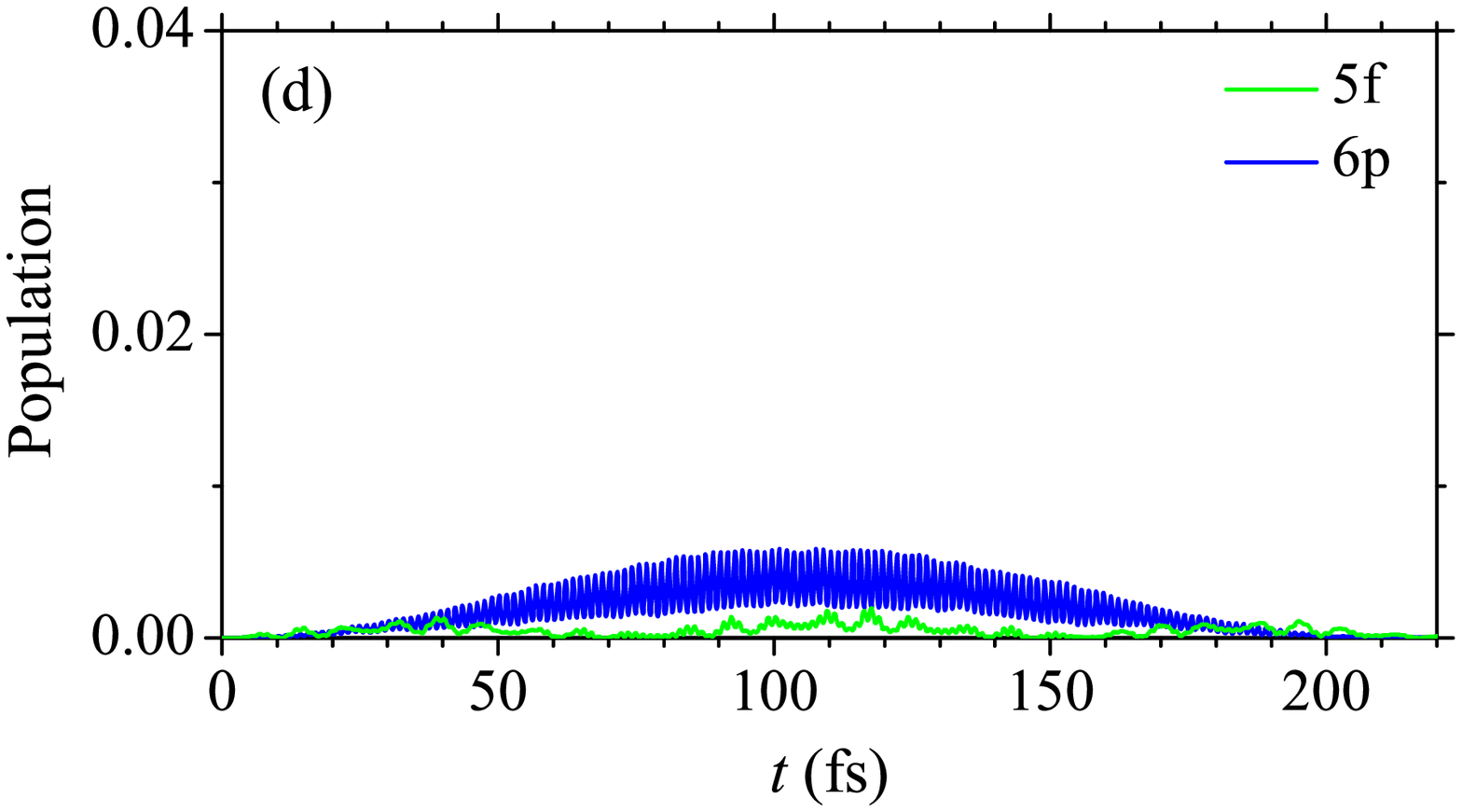}
\\
\includegraphics[width=.45\textwidth]{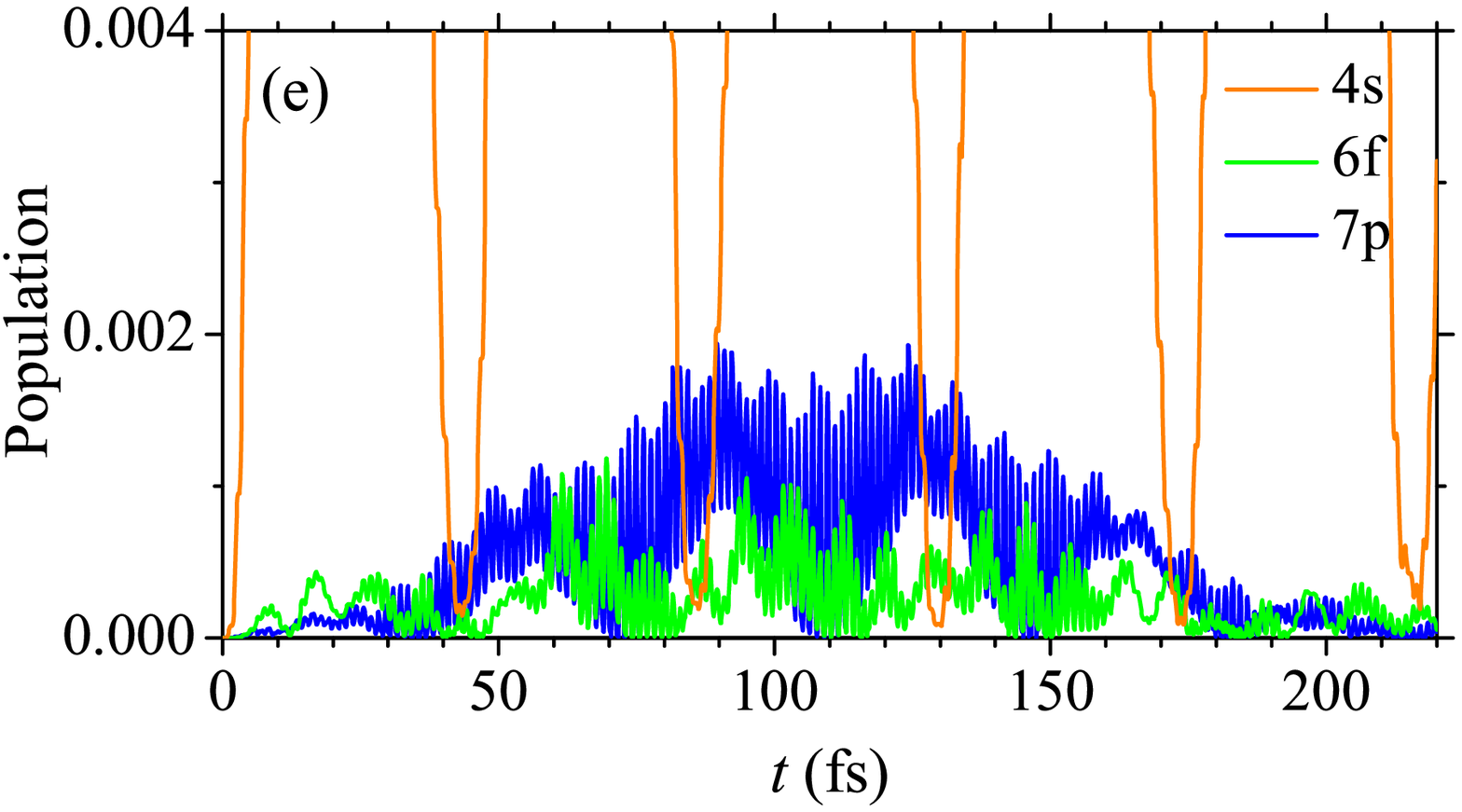}
\includegraphics[width=.45\textwidth]{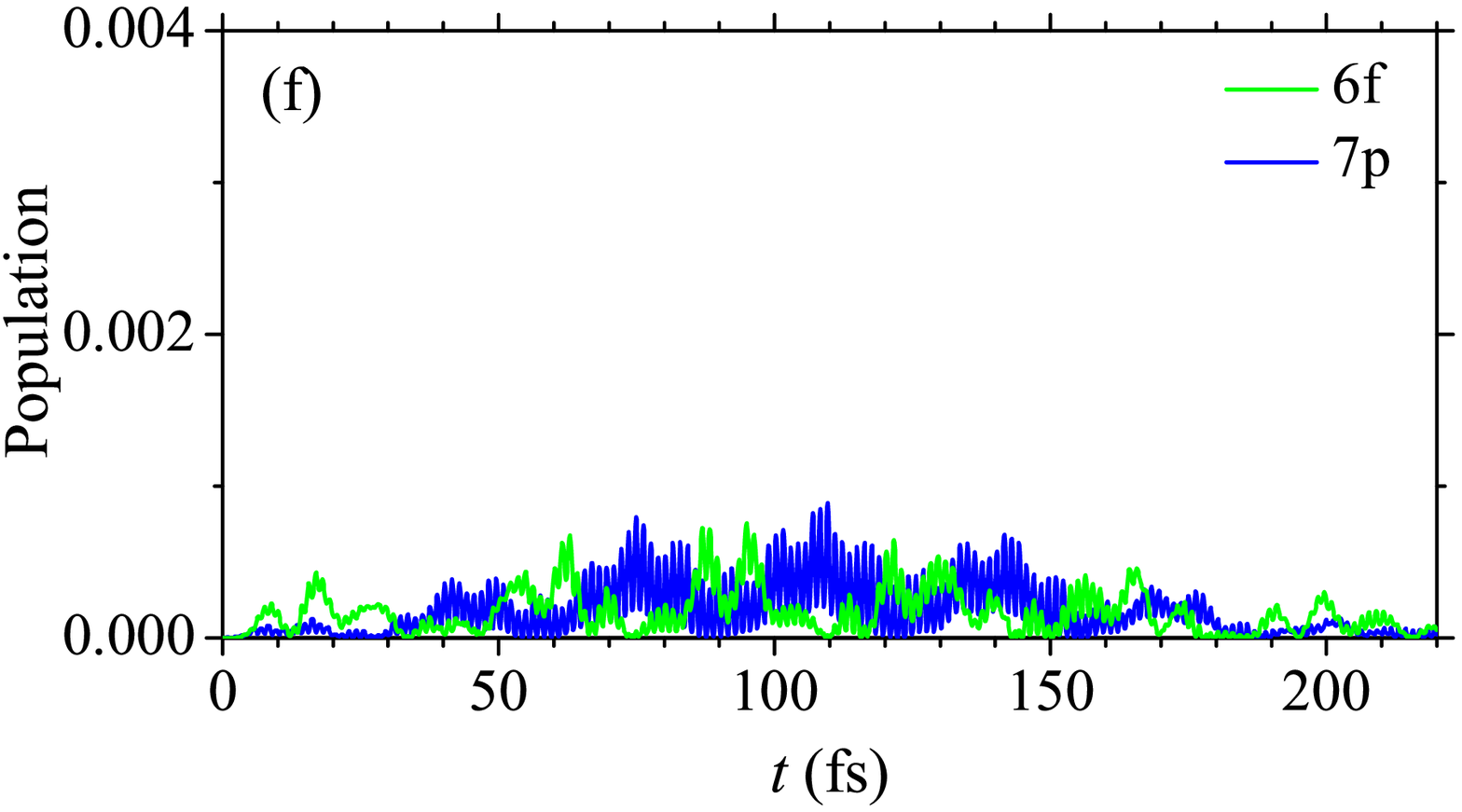}
\end{center}
\caption{Populations of states 4s, 4f, 5p, 5f, 6p, 6f and 7p of
sodium atom under the influence of continues laser field of
800\,nm wavelength and the electric field strength $F =
0.002$\,a.u., calculated by the TDC method: (left) using the basis
of 17 lowest S, P, D and F sodium states, and (right) using the
same basis, but without the 4s state.} \label{fig:pop-f.002}
\end{figure*}

\begin{figure*}
\begin{center}
\includegraphics[width=.45\textwidth]{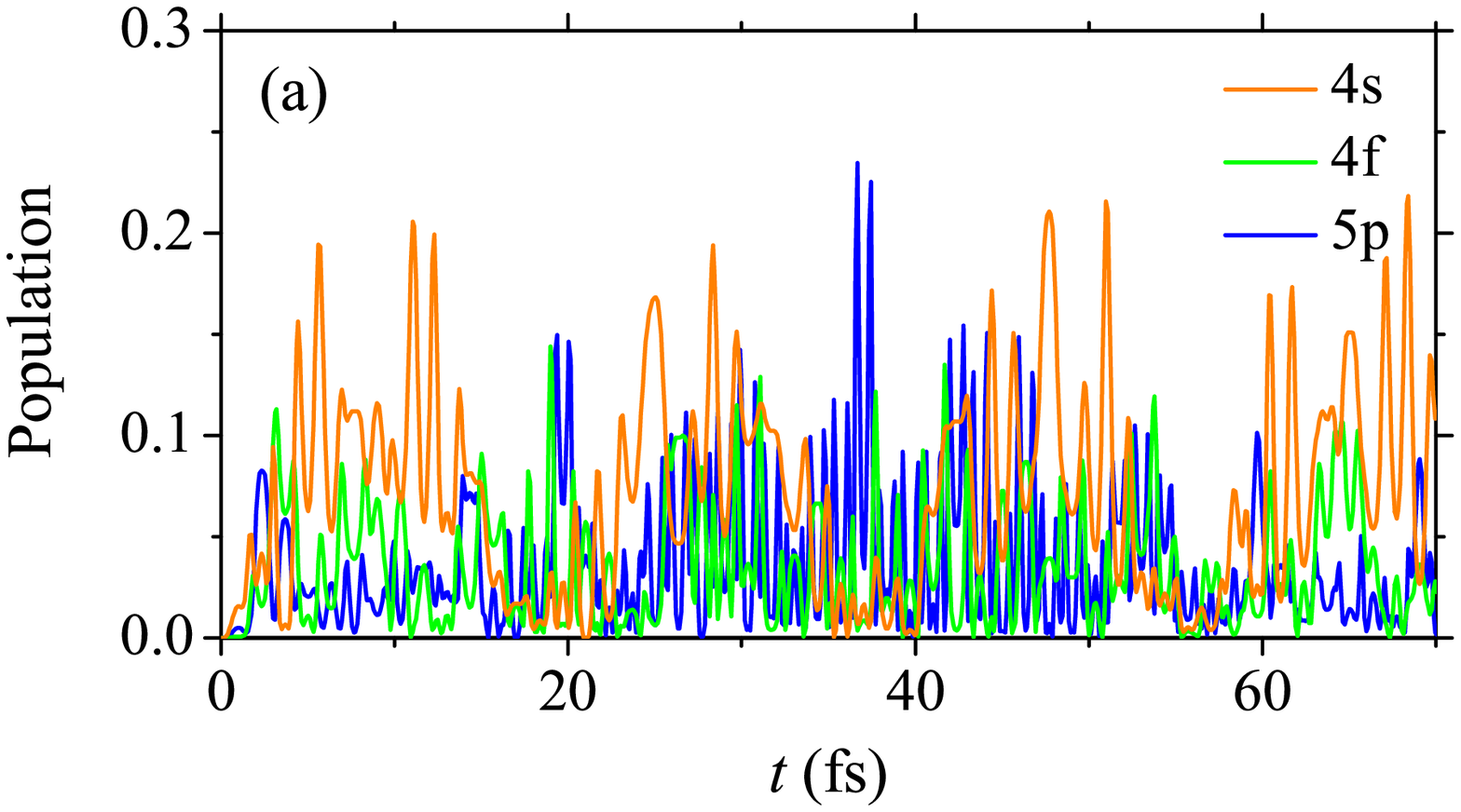}
\includegraphics[width=.45\textwidth]{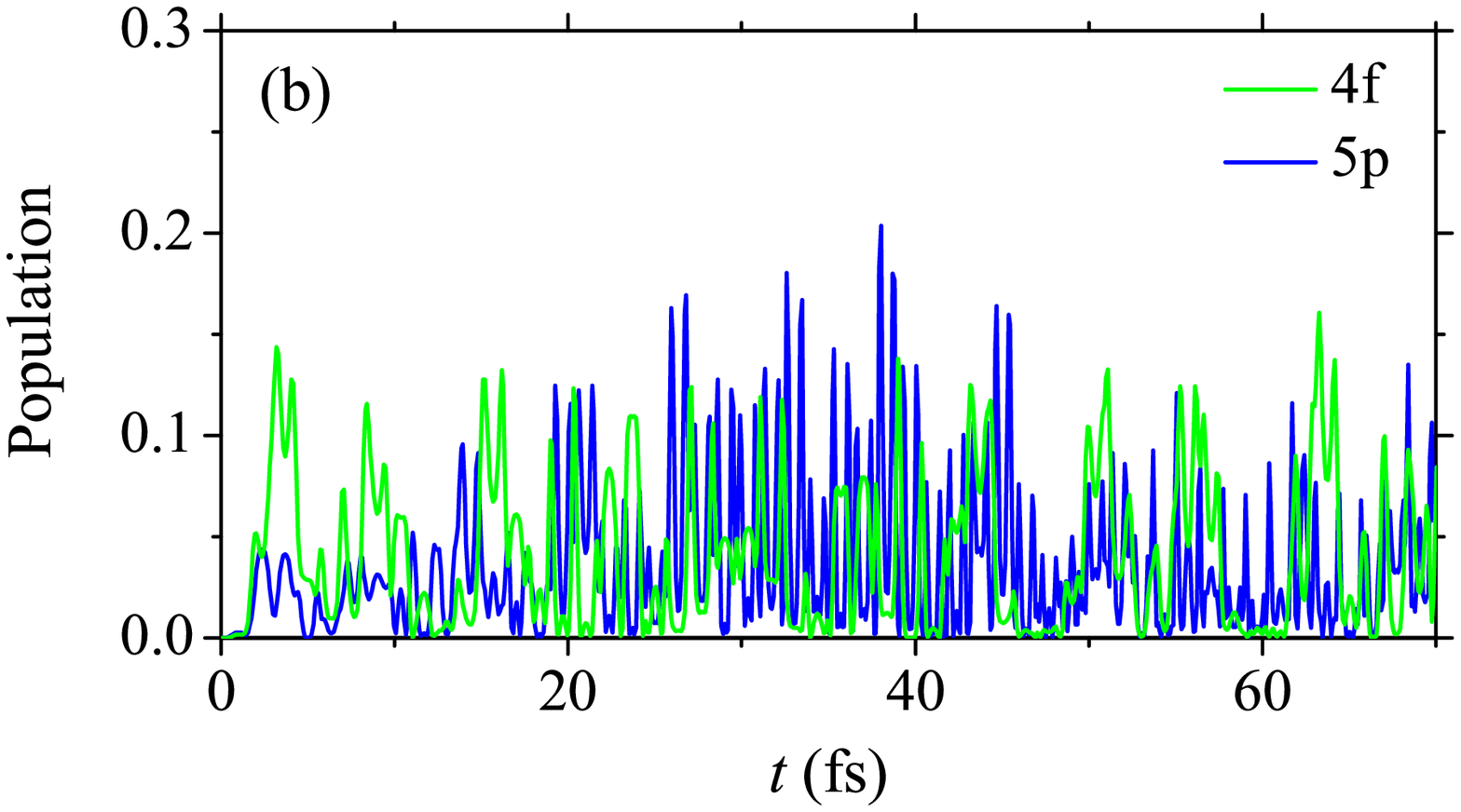}
\\
\includegraphics[width=.45\textwidth]{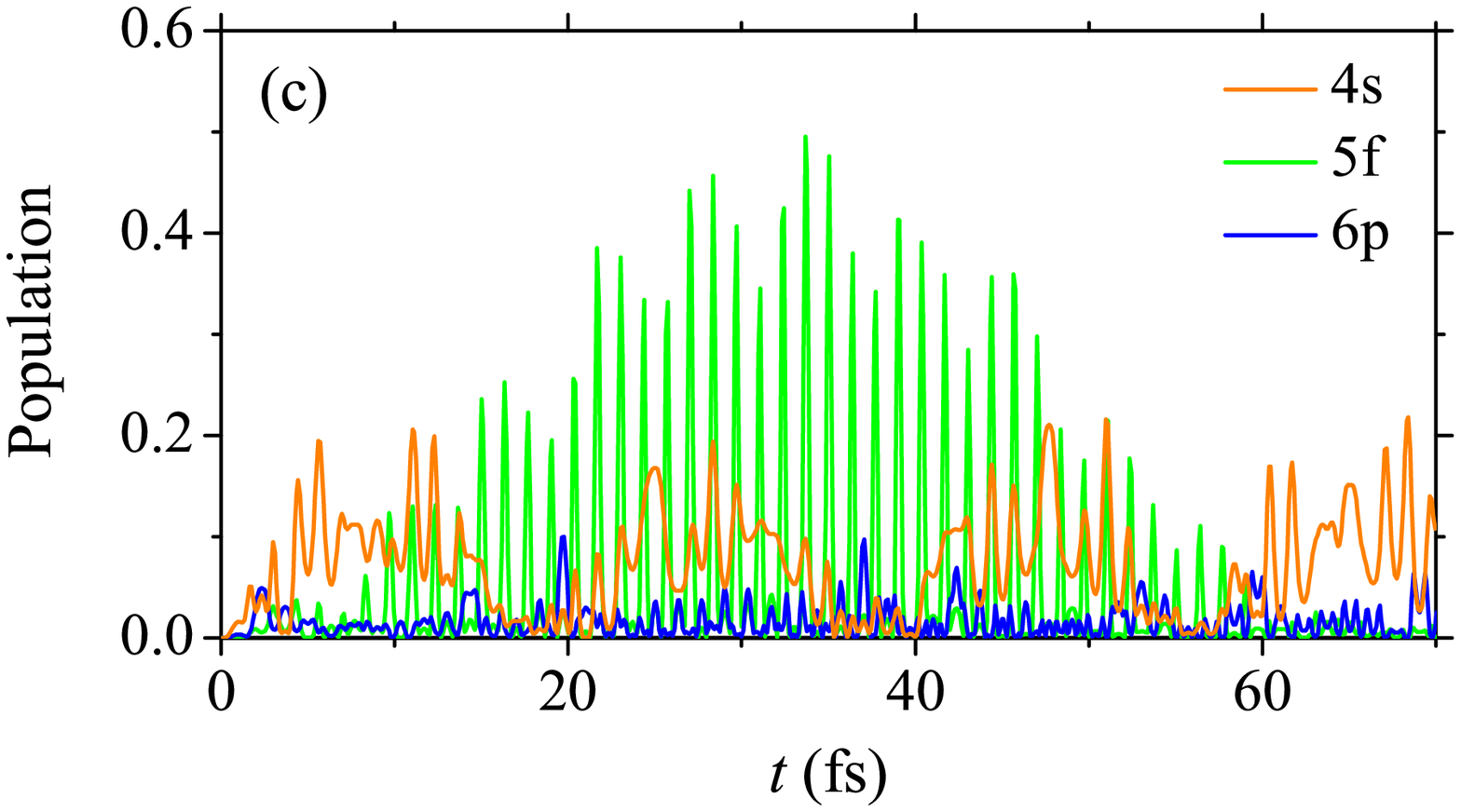}
\includegraphics[width=.45\textwidth]{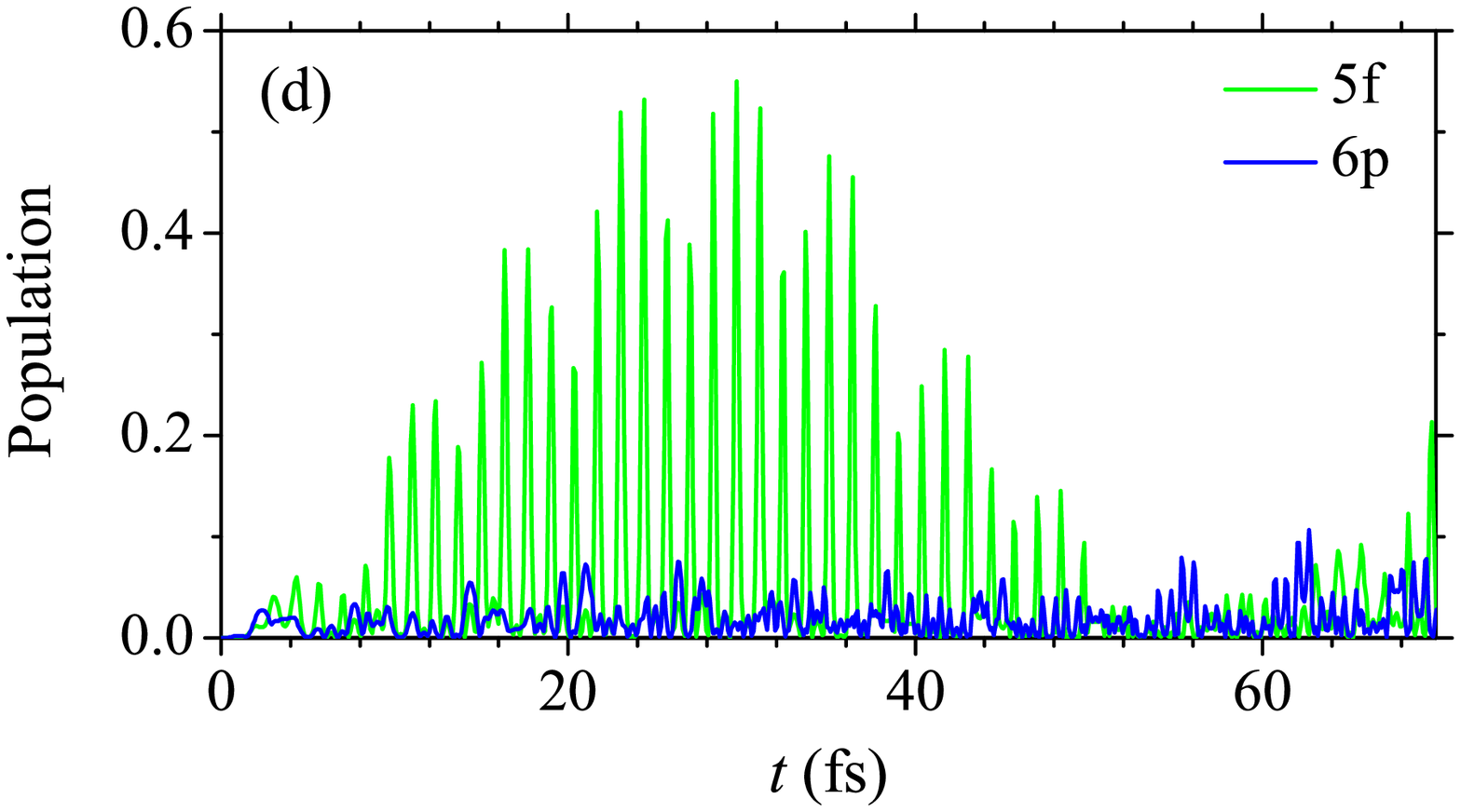}
\\
\includegraphics[width=.45\textwidth]{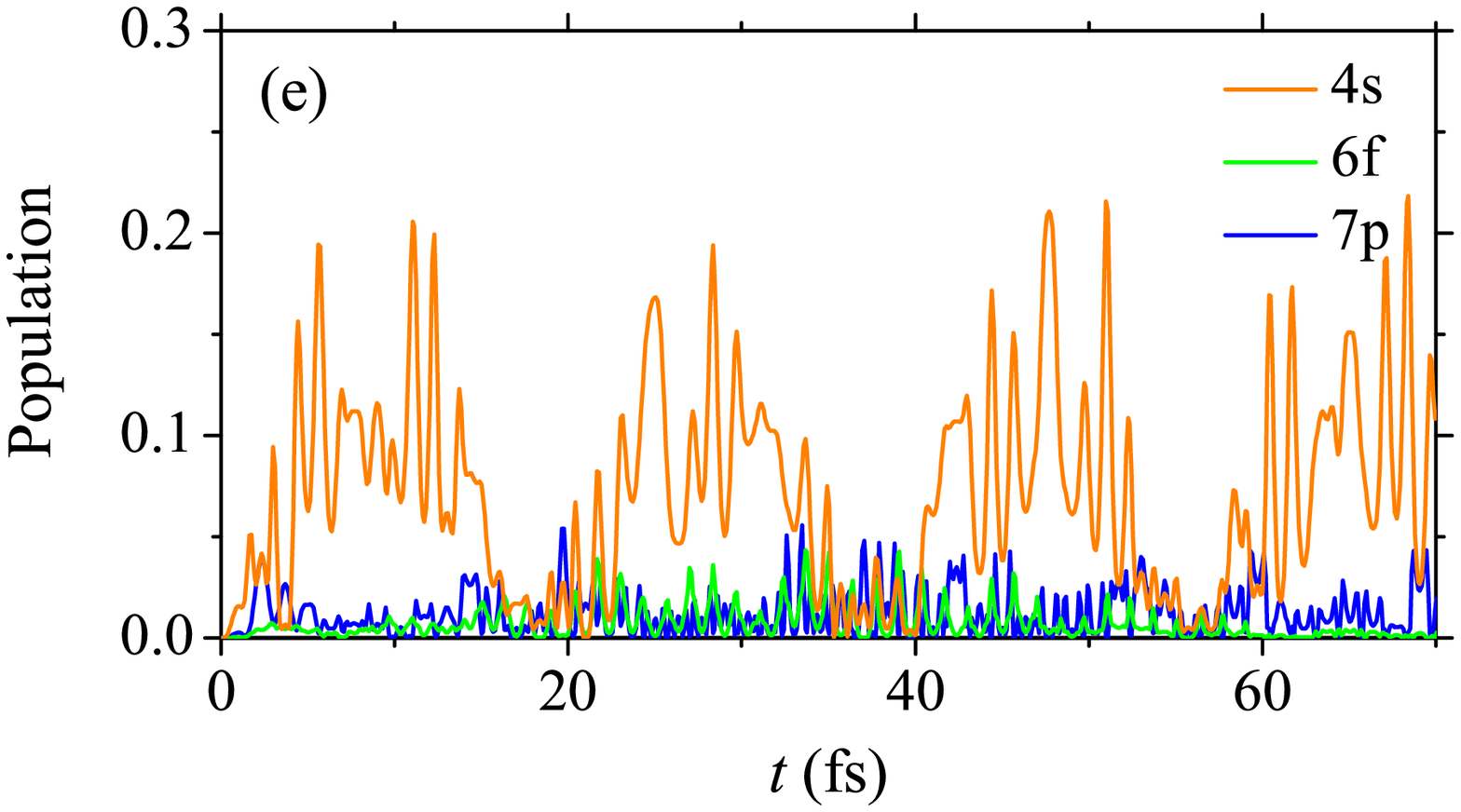}
\includegraphics[width=.45\textwidth]{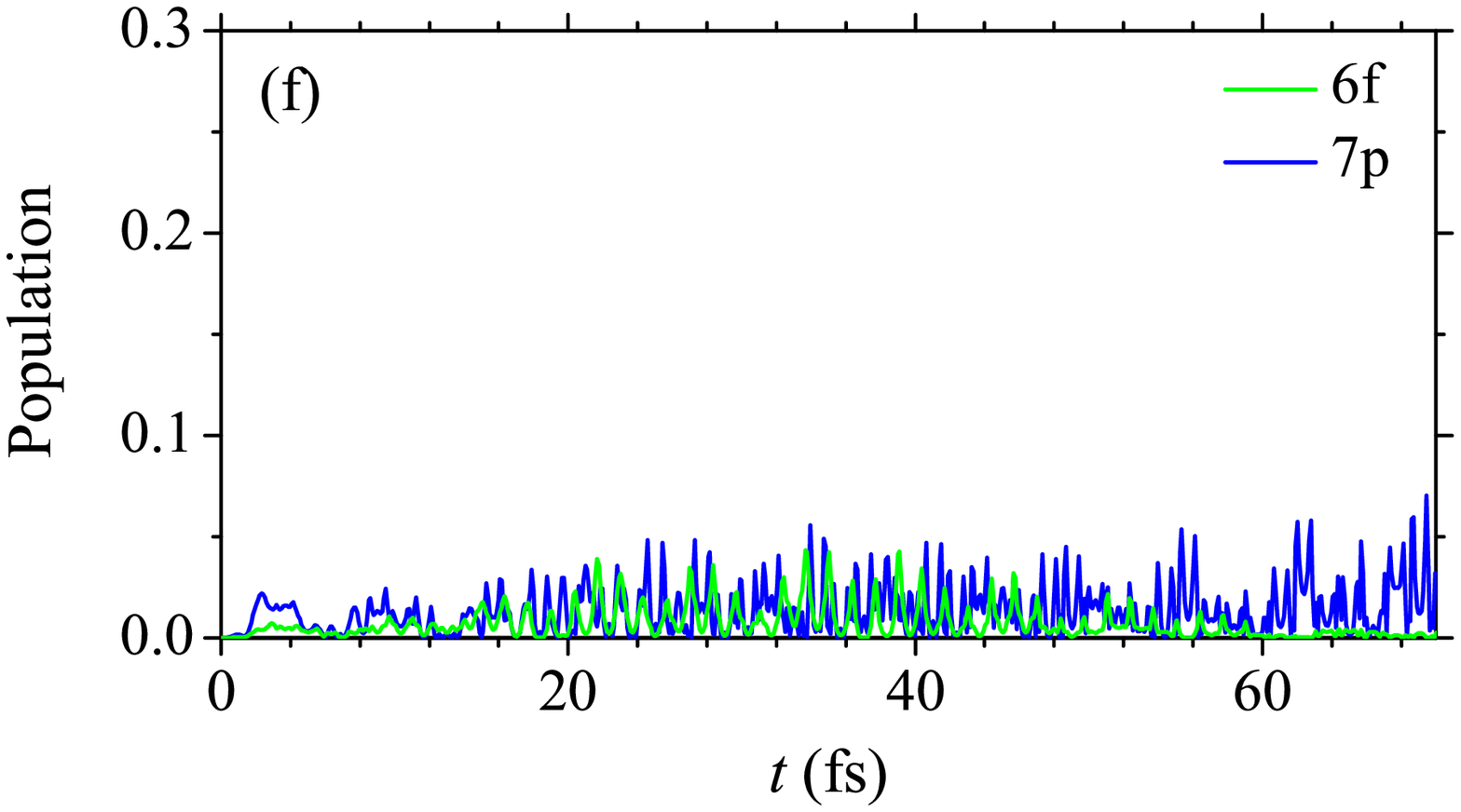}
\end{center}
\caption{The same as in Fig.~\ref{fig:pop-f.002}, but for the
field strength $F = 0.01$\,a.u.} \label{fig:pop-f.01}
\end{figure*}

Earlier experimental and theoretical studies \cite{krug,hart2016}
have indicated that a particularly important role in the MPI of
sodium using the radiation of around 800\,nm wavelength has the
2+1+1 REMPI via nearly resonant two-photon transition $3\mathrm{s}
\to 4\mathrm{s}$ and subsequent excitation of P states. Our
previous calculations \cite{pccp} have shown that this two-photon
transition is close to be resonant at the values of field strength
when the dynamic Stark shift is small (see
Fig.~\ref{fig:diagram}(c)). The TDC calculations at the field
strength $F = 0.002$\,a.u. confirm that the transfer of population
from the ground (3s) to 4s state, compared to other excited
states, is significant (see Fig.~\ref{fig:pop-f.002}). The same
calculations  at the field strength $F = 0.01$\,a.u. show,
however, that this transfer is high at stronger fields, too (see
Fig.~\ref{fig:pop-f.01}), which can be explained by strong Rabi
oscillations between these two states. The population from 4s
state flows further to state 5p (as a consequence of Rabi
oscillations between these states) and to states 6p and 7p (by the
near resonant one-photon absorption). Since $\delta E_{4s}(F)
\approx \delta E_{n\mathrm{p}}(F) \approx \delta
E_\mathrm{cb}(F)$, differences $E_{n\mathrm{p}} - E_{4s}$ and
$E_\mathrm{cb} - E_{n\mathrm{p}}$ are almost independent of $F$,
and transitions $\mathrm{4s} \to n\mathrm{p}$ and subsequent
ionization occur at all values of the field strength, producing
the photoelectrons with s and d-symmetry and excess energies given
by Eq.~(\ref{excess-e-approx2}). Therefore, the photoelectrons
produced in the 2+1+1 REMPI and in the 3+1 REMPI via the same
intermediate P state are indistinguishable in both symmetry and
energy.

An insight into relative contributions of these two ionization
channels in the total photoelectron yield can be gained {\em by
comparing} the populations of sodium P states calculated using the
model which takes into account the transitions between all
relevant states {\em with} the populations of these states
calculated using the model where the 4s state is invisible for
other states (by taking the corresponding transition matrix
elements to be zero). These calculations are done using the TDC
method where the total wave function is expanded in the basis
consisting of sodium orbitals in the SAE approximation with the
principal quantum number $n$ from 3 to 7 and the orbital quantum
number $l = 0,1,2,3$ (17 states). The populations of states 4s,
4f, 5p, 5f, 6p, 6f and 7p as functions of time in the case of
continuous radiation of 800\,nm wavelength are shown in
Figs.~\ref{fig:pop-f.002} and \ref{fig:pop-f.01} for $F =
0.002$\,a.u. and $F = 0.01$\,a.u., respectively. The plots (a),
(c), (e) are obtained taking into account all allowed transitions
between basis states, while the plots (b), (d), (f) are determined
taking into account all allowed transitions except those related
to 4s state. One can see that the presence of transition $3s \to
4s$, which is nearly resonant at $F = 0.002$\,a.u., significantly
increases the population of P states at this field strength, but
does not affect the F states. Contrarily, at $F = 0.01$\,a.u. this
transition does not affect significantly the population of higher
states. Based on this, we expect that at smaller values of the
field strength ($F < 0.004$\,a.u.) the 2+1+1 process should
dominate over the 3+1 one, but at stronger fields it might be
suppressed.

\section{Photoionization by femtosecond pulses}\label{sec:results}
\subsection{Photoelectron momentum distribution}\label{sec:pmd}

The photoelectron momentum distribution (PMD)
$|\bar\psi(\mathbf{k})|^2$ is determined from the outgoing part of
the wave function $\psi(\mathbf{r},t)$ at $t = 145\,\mathrm{fs}$
($\approx 6000\,\mathrm{a.u.} > T_p$), which was calculated by the
wave-packet propagation method (see Sec.~\ref{sec:model}). The
transformation from the coordinate to momentum representation was
done by the Fourier transform. In our case, due to the axial
symmetry of the problem, the wave function is independent of angle
$\varphi$ and it is not necessary to calculate the full 3D Fourier
transform. The PMD in the $(k_\rho,k_z)$-plane is here obtained
directly from the outgoing wave part of the function
$\psi(\rho,z)$ by transformation
\begin{equation}
\bar\psi(k_\rho,k_z) = \frac{1}{(2\pi)^2} \int_{-\infty}^\infty
\!\!dz\,e^{-ik_z z} \int_0^\infty \!\! \rho\, d\rho\, J_0(k_\rho
\rho) \psi(\rho,z). \label{ftpsi}
\end{equation}
The integral in terms of $z$-coordinate, which is a part of this
expression, was evaluated by applying the FFT algorithm. In order
to get a clear PMD, before the transformation we removed the
atomic (bound) part of the active electron wave function
$\psi(\mathbf{r},t)$ and left only the outgoing wave. It is found
that for $t > T_p$ two parts of $\psi(\mathbf{r},t)$ separate
approximately at $r = 90$\,a.u.

\begin{figure}
\begin{center}
\includegraphics[width=.45\textwidth]{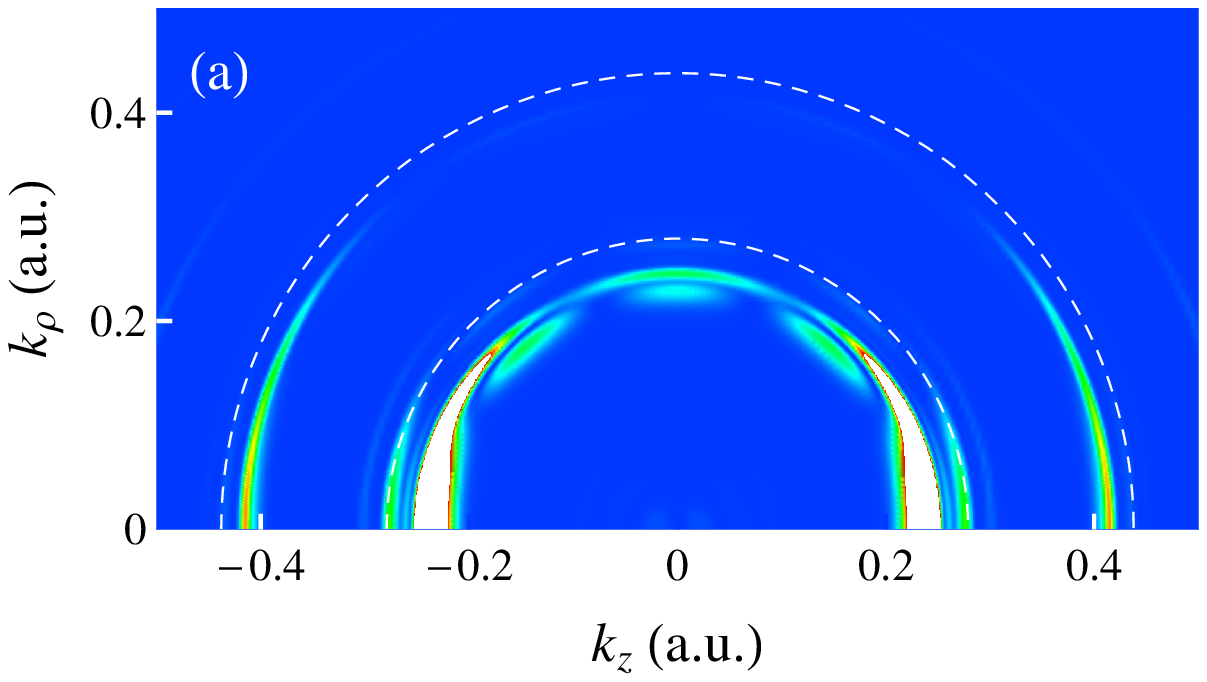}
\\
\includegraphics[width=.45\textwidth]{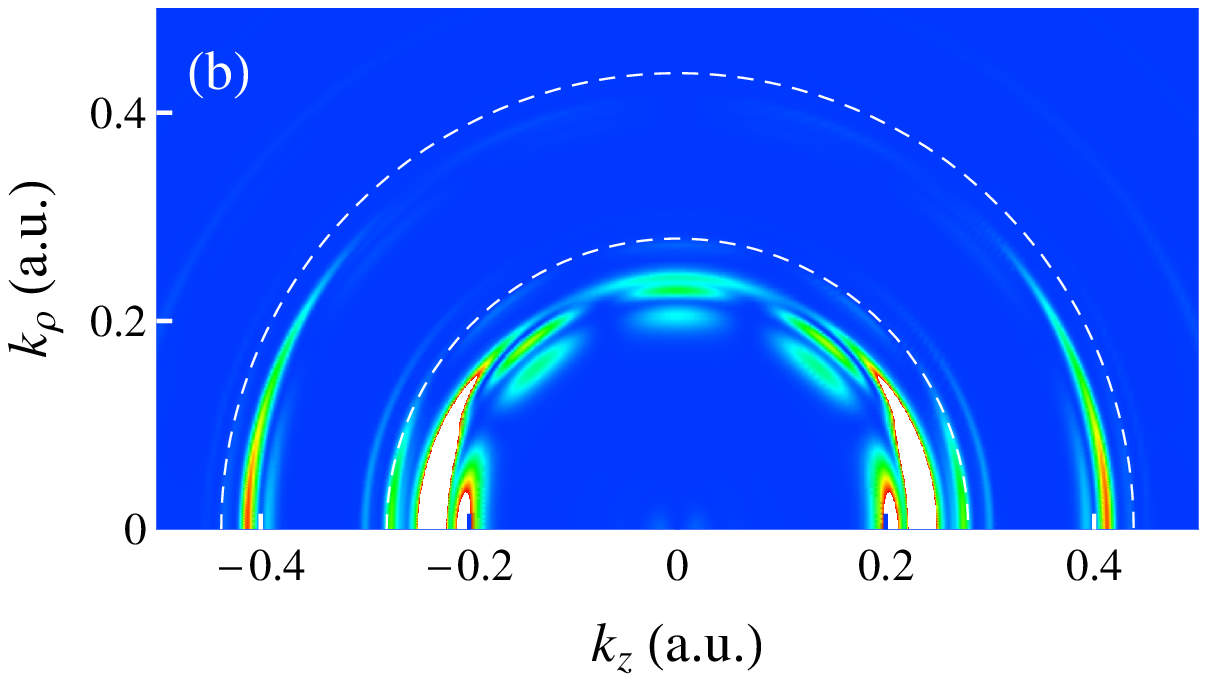}
\\
\includegraphics[width=.45\textwidth]{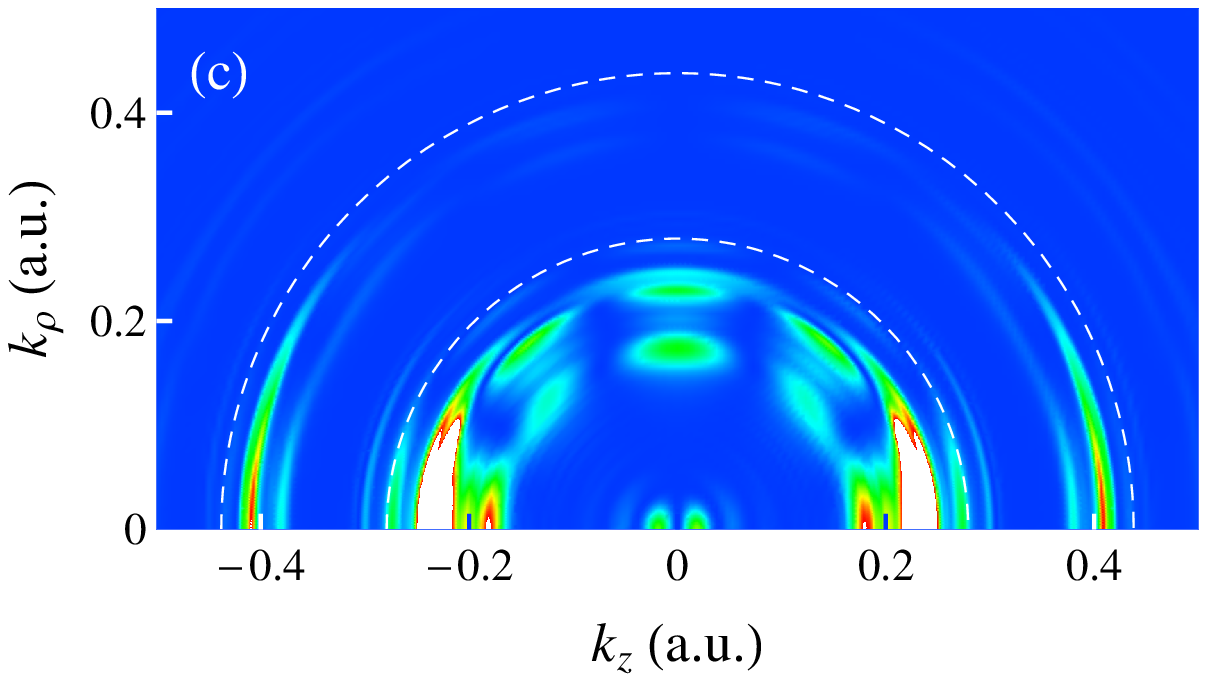}
\end{center}
\caption{Photoelectron momentum distribution
$|\bar\psi(\mathbf{k})|^2$ in the photoionization of sodium by the
laser pulse ($\lambda = 800$\,nm, $T_p = 2 \times \mathrm{FWHM} =
114$\,fs) of the form (\ref{pulse}) calculated at $t = 145$\,fs
for three values of the laser peak intensity: (a)
$3.5$\,TW/cm$^2$, (b) $4.9$\,TW/cm$^2$ and (c) $8.8$\,TW/cm$^2$.
The dashed semicircles of radii $k_0 = 0.279$\,a.u. and
$k_0^\prime = 0.438$\,a.u. correspond to the asymptotic values of
the photoelectron momentum in the weak field limit after
absorption of four and five photons (the threshold and the 1st ATI
order), respectively.} \label{fig:kdis}
\end{figure}

Fig.~\ref{fig:kdis} shows the calculated PMD for the
photoionization of sodium by 800\,nm wavelength laser pulse of the
form (\ref{pulse}) with $T_p = 114$\,fs ($\approx 4713$\,a.u.,
FWHM = 57\,fs) for three values of the peak intensity: $3.5$,
$4.9$ and $8.8$\,TW/cm$^2$ (the corresponding field strength are:
$F_\mathrm{peak} = 0.0100$, $0.0118$ and $0.0158$\,a.u.).

The radial ($k$) dependence of the PMD contains information about
the photoelectron energies ($\epsilon = \hbar^2 k^2\!/2m_{\!e}$).
The dashed semicircles of radii $k_0 = 0.279$\,a.u.
($\epsilon^{(0)} = 1.060$\,eV) and $k_0^\prime = 0.438$\,a.u.
($\epsilon^{(0)\prime} = 2.610$\,eV), drawn in the PMD plots, mark
the asymptotic values of momenta (energies) of the photoelectrons
generated in the nonresonant MPI with four and five photons,
respectively, in the weak field limit. Compared to these values,
the radial maxima of PMD determined numerically are shifted toward
the origin of $(k_\rho,k_z)$-plane. The origin of these local
maxima is twofold. Some of them are related to the nonresonant MPI
for different numbers of absorbed photons, while others can be
attributed to the REMPI (Freeman resonances). The shift of
nonresonant maxima $\delta k = \hbar^{-1}
\sqrt{2m_{\!e}\epsilon(F_\mathrm{peak})} - k_0$, referring to
Eq.~(\ref{excess-e-approx2}), is determined by the dynamic Stark
shift of the ground state and the continuum boundary at the given
laser peak intensity. The positions of Freeman resonances are, on
the other hand, almost independent on the field strength, but they
are also located below $k_0$ due to inequality $\epsilon^{(nl)}
\le \epsilon^{(0)}$ discussed in Sec.~\ref{sec:3+1REMPI}.

The angular structure of the PMD, the so-called photoelectron
angular distribution (PAD), carries information about the
superposition of accessible emitted partial waves, which,
according to selection rules for the four-photon absorption, can
be s, d and g-waves (see Fig.~\ref{fig:diagram}(a)). Indeed, apart
from the strong emission along the laser polarization direction
($\vartheta = 0^\circ$ and $180^\circ$), which can be attributed
to all three partial waves, the PADs also show maxima at
$\vartheta = 90^\circ$, which characterize d and g-waves and at
$\vartheta \approx 45^\circ$ and $135^\circ$, which characterize
the g-wave. Analogously, accessible emitted partial waves for the
five-photon absorption can be p, f and h-waves (see
Fig.~\ref{fig:diagram}(a)).

\subsection{Partial wave expansion of the outgoing wave and
photoelectron energy spectra}\label{sec:pwe-pes}

In order to determine the partial probability densities and the
PES, the outgoing wave in momentum representation is expanded in
terms of partial waves
\begin{equation}
\bar\psi(\mathbf{k}) = \sum_l \Phi_l(k) Y_{l0}(\vartheta),
\label{superpos}
\end{equation}
where $Y_{l0}(\vartheta)$ are the spherical harmonics with $m = 0$
and $\Phi_l(k) = \int Y_{l0}^*(\vartheta)\, \bar\psi(\mathbf{k})\,
\mathrm{d}\Omega$ are the corresponding radial functions. Using
the representation of $\bar\psi$ in cylindrical coordinates
determined numerically by Eq.~(\ref{ftpsi}), the radial functions
can be calculated as
\begin{eqnarray}
\Phi_l(k) = 2\pi\! \int_0^\pi \!
\bar\psi(k\sin\vartheta,k\cos\vartheta)\, Y_{l0}(\vartheta)\,
\sin\vartheta\,\mathrm{d}\vartheta. \label{radfun}
\end{eqnarray}

\begin{figure*}
\begin{center}
\includegraphics[width=.45\textwidth]{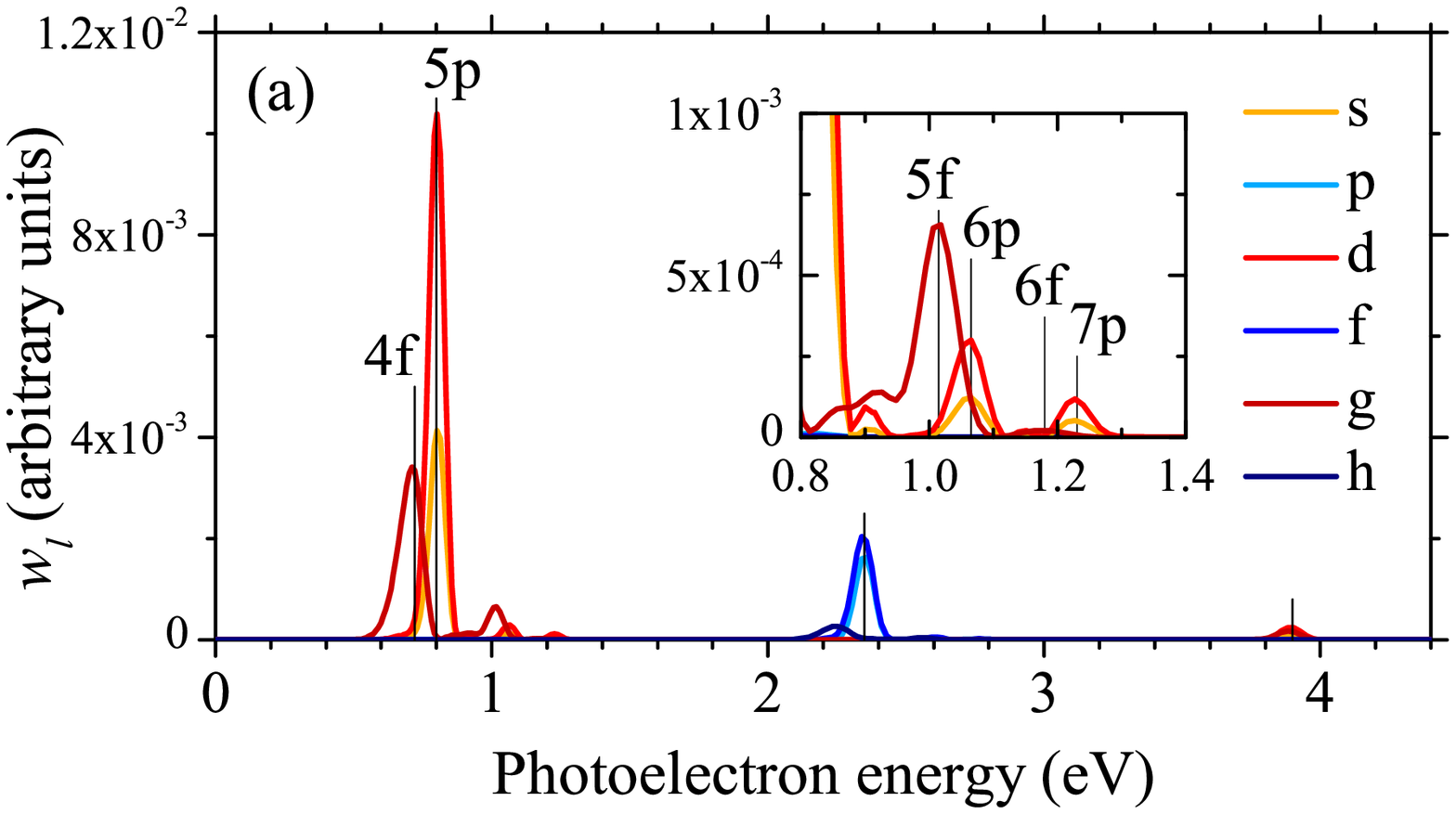}\quad
\includegraphics[width=.45\textwidth]{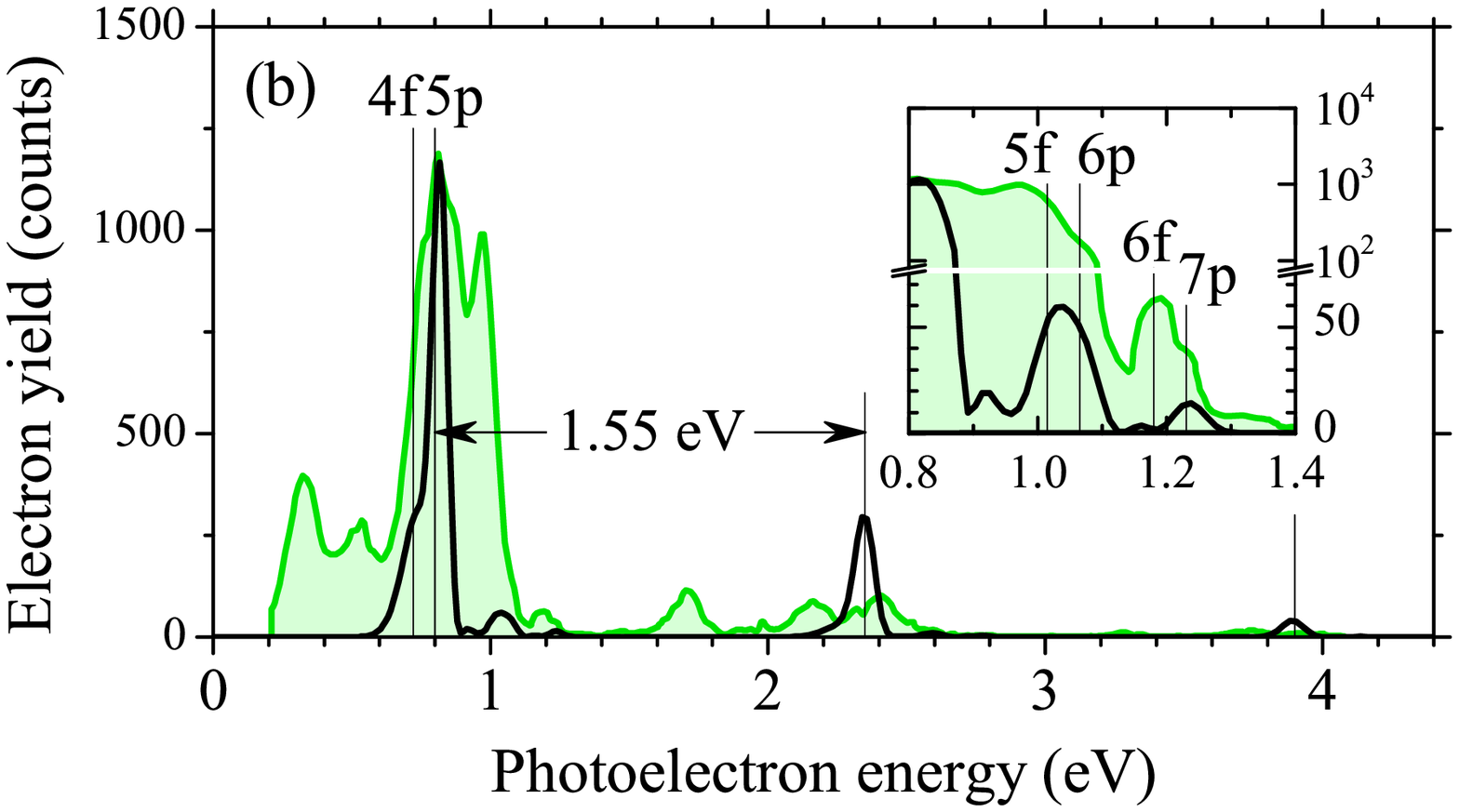}
\\
\includegraphics[width=.45\textwidth]{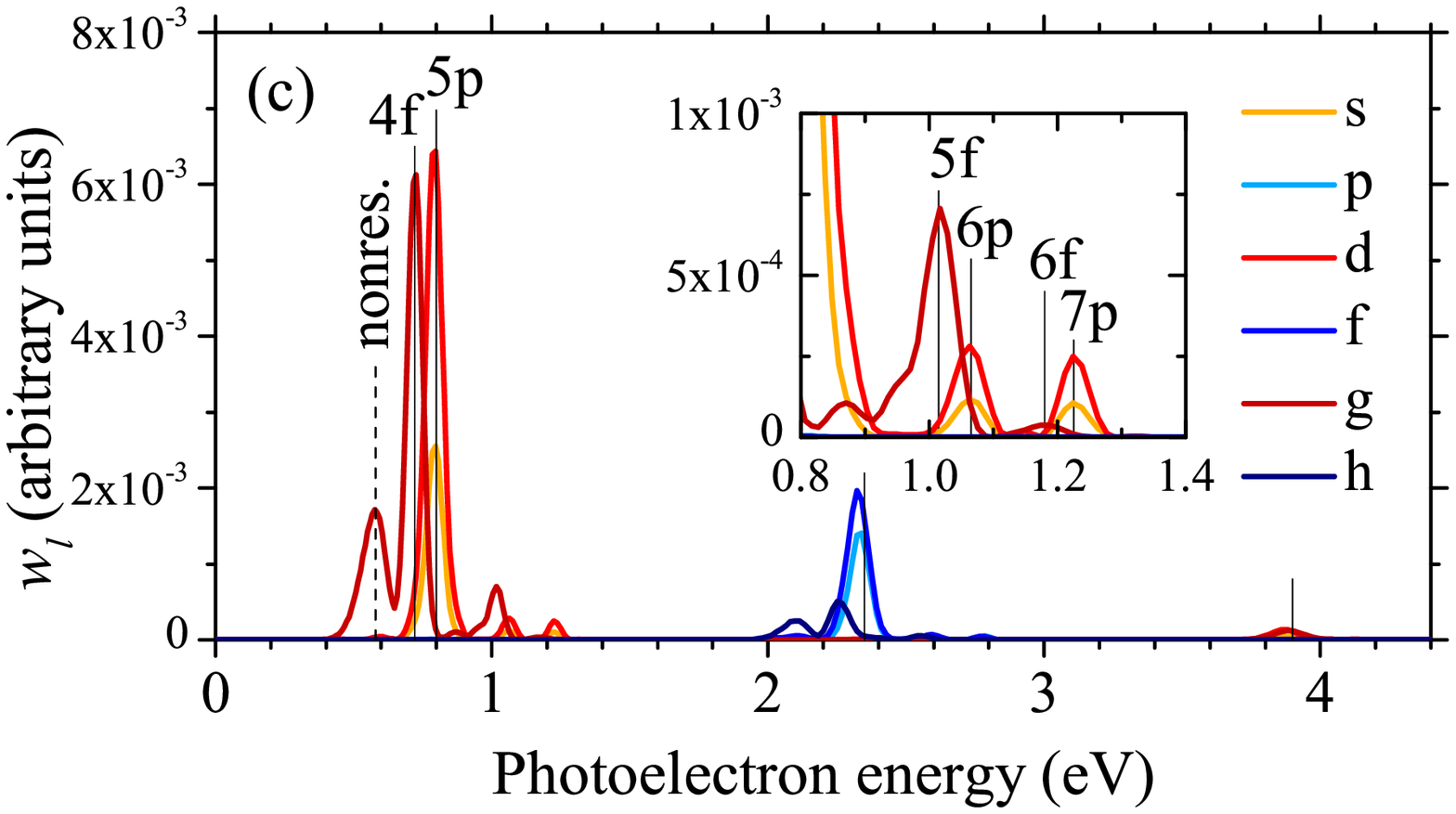}\quad
\includegraphics[width=.45\textwidth]{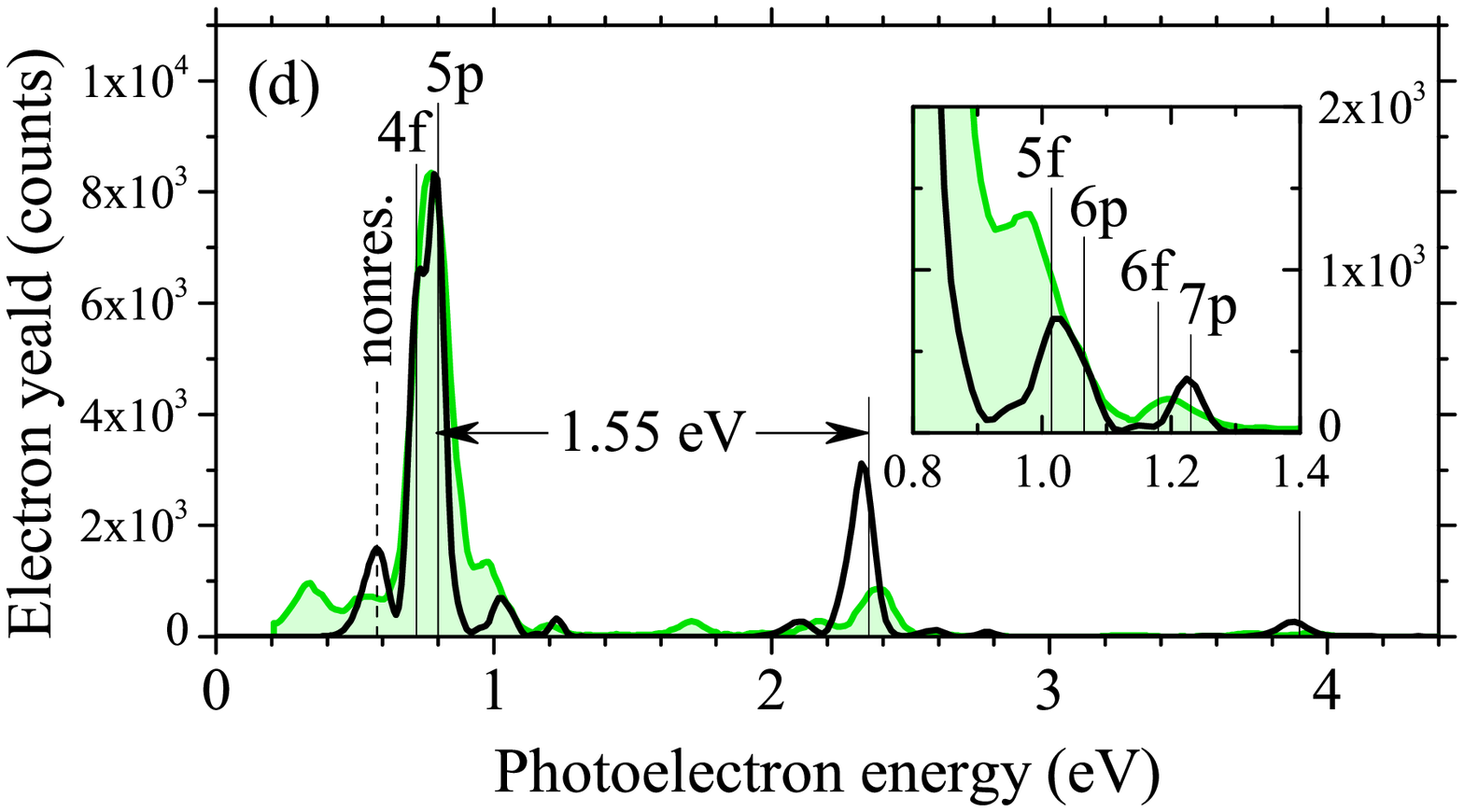}
\\
\includegraphics[width=.45\textwidth]{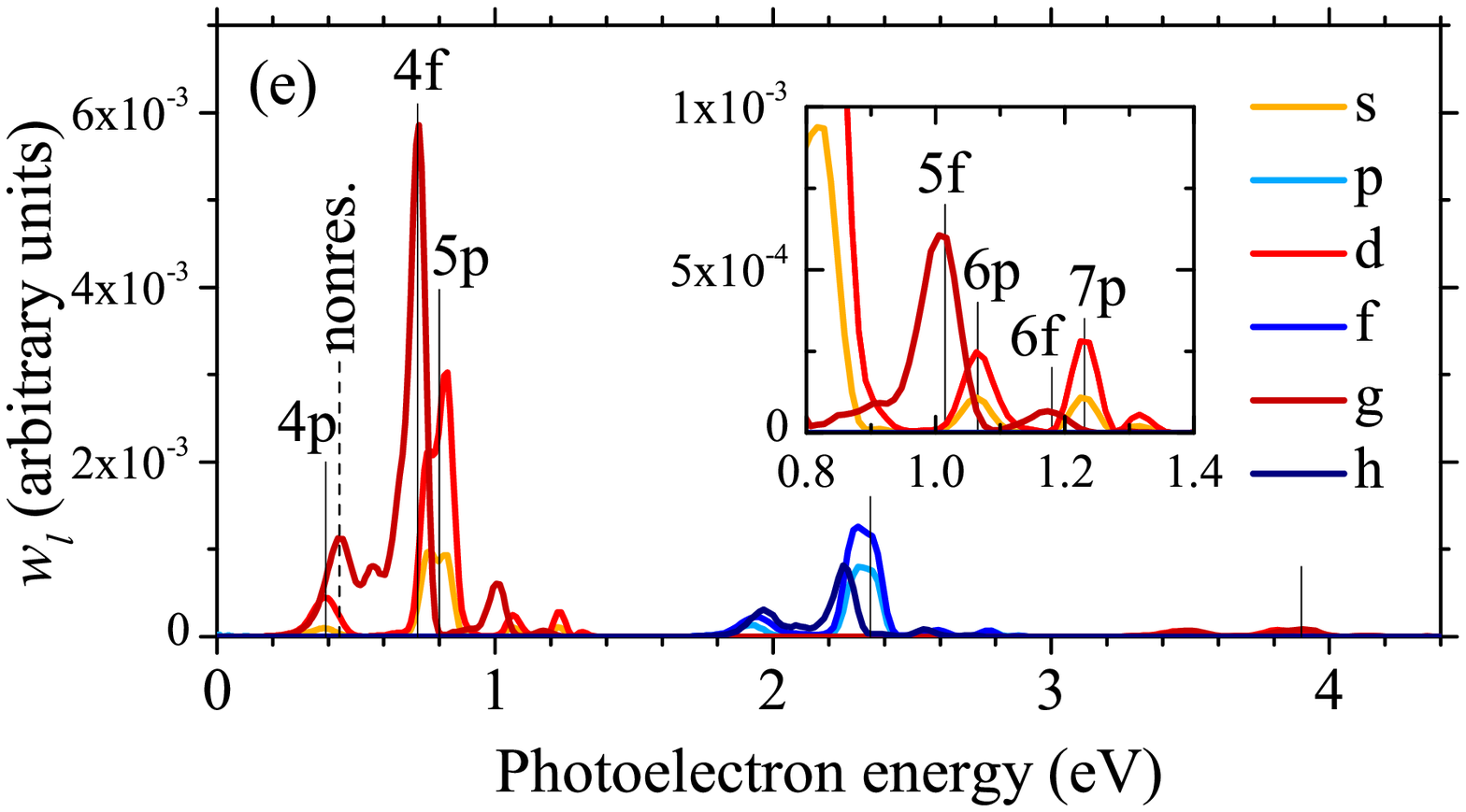}\quad
\includegraphics[width=.45\textwidth]{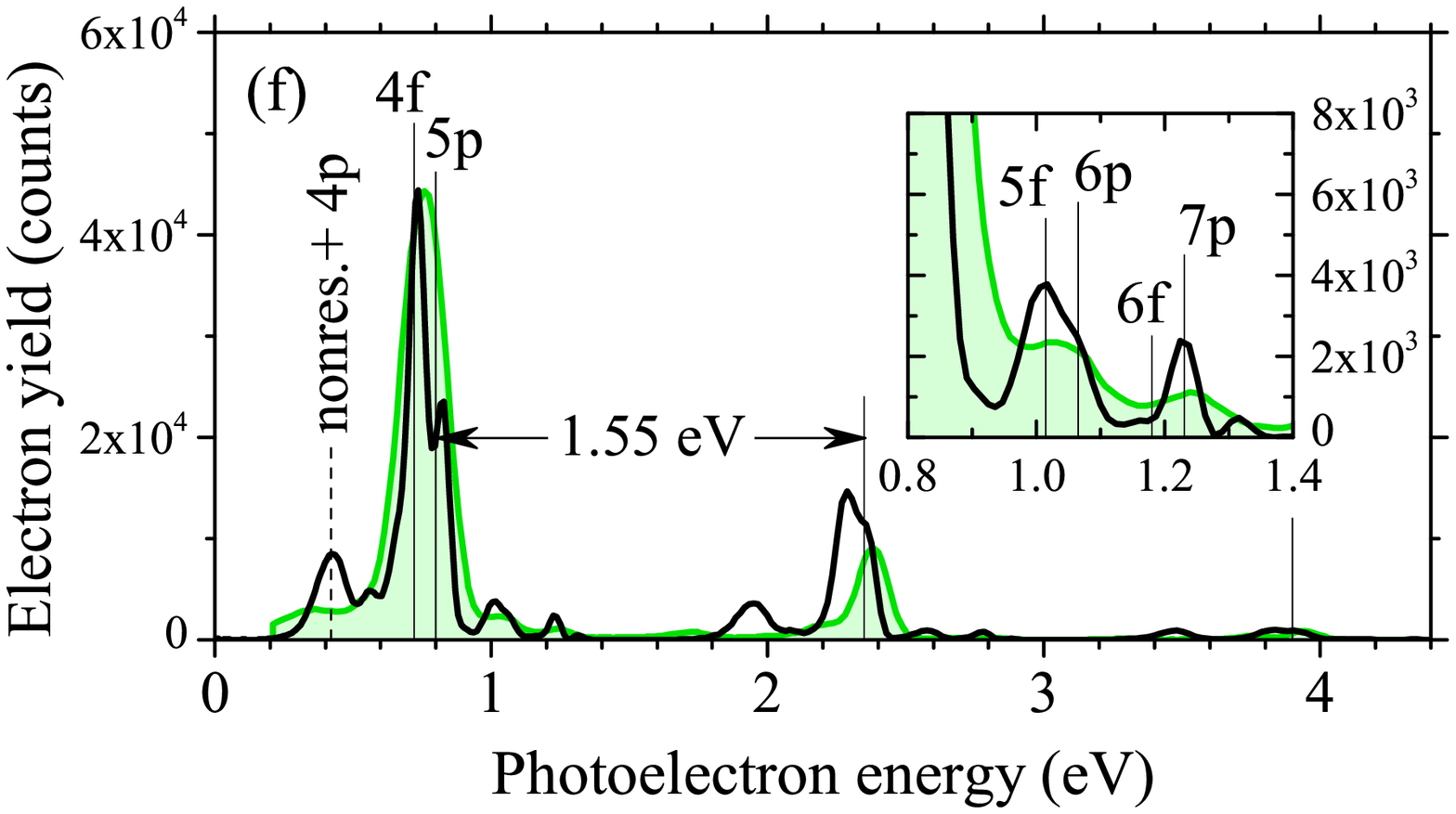}
\end{center}
\caption{Partial probability densities $w_l$ for $l = 0,\ldots,5$
(left column) and the total probability density $w$ (right column,
black line) as functions of the photoelectron energy $\epsilon =
\hbar^2k^2/(2m_e)$ obtained at three values of the laser peak
intensity: (a,b) 3.5\,TW/cm$^2$, (c,d) 4.9\,TW/cm$^2$, (e,f)
8.8\,TW/cm$^2$. Experimental results (electron yield)
\cite{hart2016} are represented by green lines (right column). The
total probability densities are rescaled for comparison with the
experimental results. The dashed vertical line indicates the
position of the nonresonant threshold (four-photon absorption)
peak, while the full vertical lines mark the energies of two REMPI
channels (via f and p states) of the threshold peak as well as the
position of 5p subpeak in the higher order ATI peaks.}
\label{fig:wl+pes}
\end{figure*}

According to partial wave expansion (\ref{superpos}), the radial
probability density of photoelectrons in momentum space is the sum
$w(k) = \sum_l w_l(k)$, where
\begin{equation}
w_l(k) = |\Phi_l(k)|^2\,k^2 \label{partw}
\end{equation}
are the partial probability densities. These quantities for $l =
0,\ldots,5$, as functions of the photoelectron excess energy
$\epsilon = \hbar^2k^2/2m_{\!e}$, are shown in the left column of
Fig.~\ref{fig:wl+pes} for three values of the laser peak
intensity: $3.5$, $4.9$ and $8.8$\,TW/cm$^2$. The corresponding
total probability densities $w$ represent the photoelectron energy
spectra (PES) for these three values of laser intensity. They are
shown in the right column of Fig.~\ref{fig:wl+pes} together with
the corresponding spectra obtained experimentally \cite{hart2016}.

The spectra, both the calculated and experimental, exhibit typical
ATI structure with prominent peaks separated by the photon energy
$\hbar\omega \approx 1.55\,\mathrm{eV}$. Fig.~\ref{fig:wl+pes}
(right column) shows the peaks corresponding to lowest three
orders of ATI (MPI by $4+s$ photons, $s = 0,1,2$) which are
located approximately at $\epsilon = 0.8\,\mathrm{eV} +
s\hbar\omega$. The partial wave analysis recovers the character of
these peaks. In the left column of Fig.~\ref{fig:wl+pes} we see
that for the photoelectron energies around the threshold peak ($s
= 0$, $\epsilon \approx 0.8$\,eV) and around the second-order ATI
peak ($s = 2$, $\epsilon \approx 3.9$\,eV) dominant contributions
in the total probability density come from the partial waves with
even $l$ (s, d, g-waves). Thus, the photoelectrons with these
energies are generated by absorbing an even number of photons ($N
= 4$ and 6). Contrarily, in the vicinity of the first-order ATI
peak ($s = 1$, $\epsilon \approx 2.35$\,eV) the partial waves with
even $l$ are suppressed and those with odd $l$ (p, f, h-waves)
dominate. Therefore, in this case odd number of photons is
absorbed (here $N = 5$). Each ATI peak, in addition, has an
internal structure in the form of local (sub)peaks which can be
attributed to the nonresonant MPI and to the REMPI via different
excited states.

\subsection{Nonresonant photoionization}\label{sec:nores}

The position of the nonresonant threshold peak (four-photon
ionization maximum) predicted by formula (\ref{excess-e-approx2})
for laser peak intensities 3.5, 4.9 and 8.8\,TW/cm$^2$ is
$\epsilon(F_\mathrm{peak}) = 0.74$\,eV, 0.61\,eV and 0.26\,eV,
respectively. This peak is visible in the numerically determined
spectra shown in Fig.~\ref{fig:wl+pes}. Since the energy of
photoelectrons produced by the nonresonant MPI does not depend on
$l$, a feature of the nonresonant peak is that the maxima of
contributing partial densities $w_l$ have the same positions on
the energy axis. At the laser peak intensity of $3.5$\,TW/cm$^2$,
however, the nonresonant peak overlaps with the most prominent
REMPI peak (see Fig.~\ref{fig:wl+pes}(a,b)) and it is difficult to
estimate the position of former from the numerical data. The
position of this peak at intensities $4.9$\,TW/cm$^2$ and
$8.8$\,TW/cm$^2$ is 0.58\,eV and 0.44\,eV (numerical values),
respectively (see Figs.~\ref{fig:wl+pes}(c-f)). A discrepancy
between the values obtained by formula (\ref{excess-e-approx2})
and from numerical calculations (especially at $I =
8.8$\,TW/cm$^2$) is attributed to the approximative character of
the former. In addition, it should be mentioned that in
experimental spectra the nonresonant peak is less prominent
(almost invisible). This observation is reported also in an
earlier work presenting a comparison between calculated and
experimental data for the photoionization of lithium
\cite{morishita}. Nonresonant peaks of the first and of the second
ATI order can be observed in Fig.~\ref{fig:wl+pes}, too, at
positions which are shifted by one and two-photon energy relative
to the threshold peak at a given laser intensity.

\subsection{Resonantly enhanced multiphoton ionization}\label{sec:rempi}

In contrast to nonresonant peaks the positions of REMPI peaks
(Freeman resonances), as explained in Sec.~\ref{sec:scheme}, are
almost independent of the laser peak intensity. We anticipated in
that section that photoelectrons belonging to the threshold peak
(four-photon ionization) reach the continuum along tree pathways
which involve: (i) the 3+1 REMPI via intermediate P states, (ii)
the 3+1 REMPI via intermediate F states and (iii) the 2+1+1
process via nearly resonant transition $3\mathrm{s} \to
4\mathrm{s}$ and subsequent excitation of P-states. For the most
prominent peak at $\epsilon \approx 0.8$\,eV the intermediate P
and F states are the states 5p and 4f, whereas for the subpeaks at
$\epsilon \approx 1$\,eV and at $\epsilon \approx 1.2$\,eV these
are the states 6p and 5f and the states 7p and 6f, respectively.
Note, however, that for the radiation of 800\,nm wavelength the
transfer of population from the ground state to states 7p and 6f
is only near resonant ($E_\mathrm{7p}, E_\mathrm{6f} >
3\hbar\omega - I_p$, see Fig.~\ref{fig:diagram}(b)) and, strictly
speaking, the four-photon ionization via these states is not 3+1
REMPI (see Sec.~\ref{sec:3+1REMPI}). In this case formula
(\ref{excess-e-approx1}) is not applicable, but the photoelectron
energy can be estimated using relation (\ref{excess-e-approx2}).

Here we focus on the main subpeak of the threshold peak (around
0.8\,eV). Taking into account all three ionization pathways (see
the first paragraph in Sec.~\ref{sec:scheme}) and the fact that
photoelectrons produced in the 3+1 and 2+1+1 REMPI via P states
cannot be distinguished (Sec.~\ref{sec:2+1+1REMPI}), the electron
outgoing wave in the energy domain of this peak can be written as
the superposition of two wave-packets
\begin{equation}
\bar\psi = \bar\psi^\mathrm{(5p)} + \bar\psi^\mathrm{(4f)},
\label{wp5p+wp4f}
\end{equation}
which, according to Fig.~\ref{fig:diagram}(a), have forms
\begin{eqnarray}
\bar\psi^\mathrm{(5p)} &=& \Phi_0^\mathrm{(5p)} Y_{00} +
\Phi_2^\mathrm{(5p)} Y_{20}, \label{wp5p}
\\[.5ex]
\bar\psi^\mathrm{(4f)} &=& \Phi_2^\mathrm{(4f)} Y_{20} +
\Phi_4^\mathrm{(4f)} Y_{40}. \label{wp4f}
\end{eqnarray}
Since states 5p and 4f shift into the three-photon resonance at
different field strengths (see Table~\ref{table1} and
Fig.~\ref{fig:diagram}(b)), wave packets (\ref{wp5p}) and
(\ref{wp4f}) are formed in different phases of the laser pulse and
characterized by different mean energies ($\approx 0.8$\,eV and
0.7\,eV, respectively, referring to Table~\ref{table1}).

Expression (\ref{wp5p+wp4f}) with components (\ref{wp5p}),
(\ref{wp4f}) is compatible with the partial wave expansion of
function $\bar\psi$. The left column of Fig.~\ref{fig:wl+pes}
demonstrates that the outgoing wave in the domain of threshold
peak decomposes into s, d and g-waves
\begin{equation}
\bar\psi = \Phi_0 Y_{00} + \Phi_2 Y_{20} + \Phi_4 Y_{40}.
\label{decomp}
\end{equation}
Radial functions $\Phi_l$ and the corresponding partial
probability densities $w_l$ are determined numerically using
formulae (\ref{radfun}) and (\ref{partw}), respectively. The
positions of maxima of $w_l(\epsilon)$ (see Table~\ref{table2})
confirm the existence of two electron wave-packets with different
mean energies. The photoelectrons with s and d-symmetry have a
higher mean energy ($\approx 0.8$\,eV) than those with g-symmetry
(around $0.7$\,eV). Referring to Table~\ref{table1}, these two
energies characterize the photoelectrons produced in the REMPI via
5p and 4f states ($\epsilon^\mathrm{(5p)}$ and
$\epsilon^\mathrm{(4f)}$). Since the maxima of $w_2(\epsilon)$ and
$w_0(\epsilon)$ almost coincide, we conclude that s-electrons and
the majority of d-electrons belong to the same channel, i.e. they
are generated in the 3+1 (or 2+1+1) REMPI via 5p state. Thus, the
contribution of d-electrons in the wave packet (\ref{wp4f}) is
minor ($\Phi_2^\mathrm{(4f)} \approx 0$). Then, comparing
expansion (\ref{decomp}) and expressions (\ref{wp5p+wp4f}),
(\ref{wp5p}), (\ref{wp4f}), we get $\Phi_0^\mathrm{(5p)} =
\Phi_0$, $\Phi_2^\mathrm{(5p)} \approx \Phi_2$ and
$\Phi_4^\mathrm{(4f)} = \Phi_4$.

\begin{table}
{\small \caption{Photoelectron energies $\epsilon$ at which the
partial probability densities $w_l$ ($l = 0,2,4$) shown in
Fig.~\ref{fig:wl+pes}(a,c,e) take maximal values $w_l^\mathrm{m}$
and the ratios $w_l^\mathrm{m}/w_0^\mathrm{m}$ ($l = 2,4$).}
  \label{table2}
  \begin{indented}
  \item[]
  \begin{tabular}{@{}cccccc}
  \br
    partial wave: & s ($l = 0$) & \multicolumn{2}{c}{d ($l = 2$)} & \multicolumn{2}{c}{g ($l =
    4$)} \\
    \mr
    $I$ (TW/cm$^2$) & $\epsilon$\,(eV)
                    & $\epsilon$\,(eV) & $w_2^\mathrm{m}/w_0^\mathrm{m}$
                    & $\epsilon$\,(eV) & $w_4^\mathrm{m}/w_0^\mathrm{m}$ \\
    \mr
    3.5 & 0.80 & 0.80 & 2.48 & 0.71 & 0.81 \\
    4.9 & 0.79 & 0.79 & 2.50 & 0.72 & 2.38 \\
    8.8 & 0.79 & 0.79 & 3.05 & 0.72 & 5.90 \\
    \br
  \end{tabular}
  \end{indented}
}
\end{table}

A similar analysis indicates that the subpeak at $\epsilon \approx
1$\,eV is related to 3+1 REMPI via states 5f and 6p, while that at
$\epsilon \approx 1.2$\,eV is related to 3+1 (or 2+2+1) REMPI via
7p state (or sequence $4\mathrm{s} \to 7\mathrm{p}$) and much less
to 3+1 REMPI via state 6f. Finally, note that for 8.8\,TW/cm$^2$
laser intensity an additional peak arises at $\epsilon \approx
0.39$\,eV (see Fig.~\ref{fig:wl+pes}(e)). It can be related to 3+1
REMPI via state 4p \cite{pccp}. In the PES this peak overlaps with
the nonresonant peak at $\epsilon \approx 0.44$\,eV (see
Fig.~\ref{fig:wl+pes}(f)).

\subsection{Selective enhancement of photoionization channels}
\label{sec:select}

Partial-wave analysis of the photoelectron wave function (outgoing
wave) revealed that each subpeak of the threshold peak and
succeeding ATI peaks, shown in Fig.~\ref{fig:wl+pes}(b,d,f),
contains contributions of three REMPI channels, i.e. it is a
superposition of three Freeman resonances. Here we examine their
relative share in the electron yield and the possibility for
selective ionization of atom through a single channel.

At the laser peak intensity $I = 3.5\,\mathrm{TW/cm}^2$ the
maximum contribution in the peak around 0.8\,eV comes from the
electrons of d-symmetry. At $I = 4.9\,\mathrm{TW/cm}^2$ the
electrons of d and g-symmetry have approximately equal
contributions to this peak, while at intensity $I =
8.8\,\mathrm{TW/cm}^2$ maximum contribution comes from the
electrons of g-symmetry (see the ratios of $w_l$ maxima shown in
Table~\ref{table2}). The contribution of s-electrons in this peak
is 2.5-3 times smaller than the contribution of d-electrons. The
contributions of s, d and g-electrons in the peaks at $\epsilon
\approx 1$\,eV and $\epsilon \approx 1.2$\,eV do not change
significantly with the laser intensity. The peak at $\epsilon
\approx 1$\,eV is built predominantly by the electrons of
g-symmetry, while the maximum contribution to the peak at
$\epsilon \approx 1.2$\,eV comes from the electrons of d-symmetry
(see the insets in Fig.~\ref{fig:wl+pes}(c,e)). These results
indicate that for $F \le F_\mathrm{OBI}$ dominant ionization
channel in the peak around 0.8\,eV is the 3+1 or, more likely,
2+1+1 REMPI via 5p state, while for $F > F_\mathrm{OBI}$ it is the
3+1 REMPI via 4f state. For the peaks around 1\,eV and 1.2\,eV,
dominant channels are the 3+1 REMPI via 5f state and the 2+1+1
REMPI via 7p state, respectively.

\begin{figure*}
\begin{center}
\includegraphics[width=.45\textwidth]{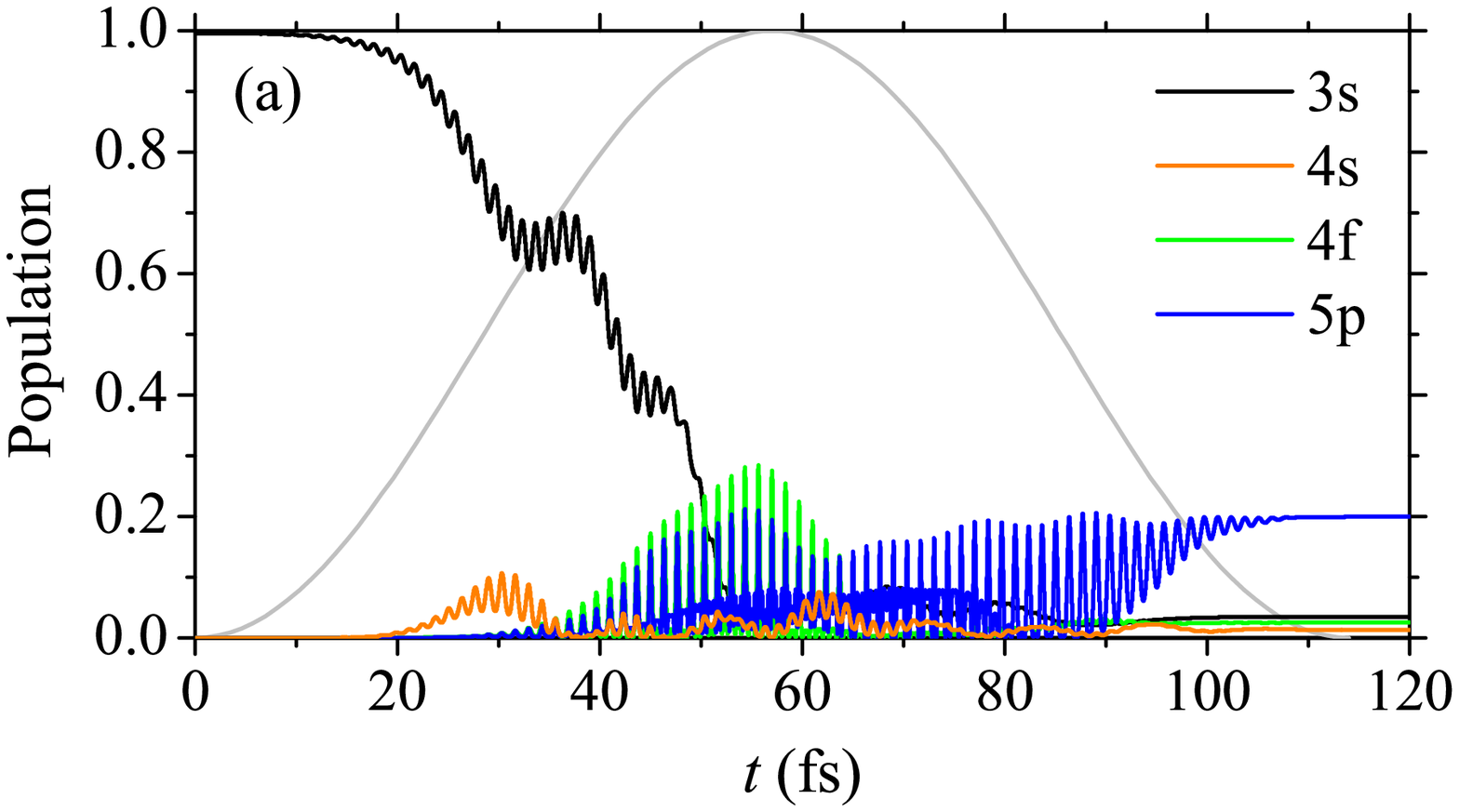}\quad
\includegraphics[width=.45\textwidth]{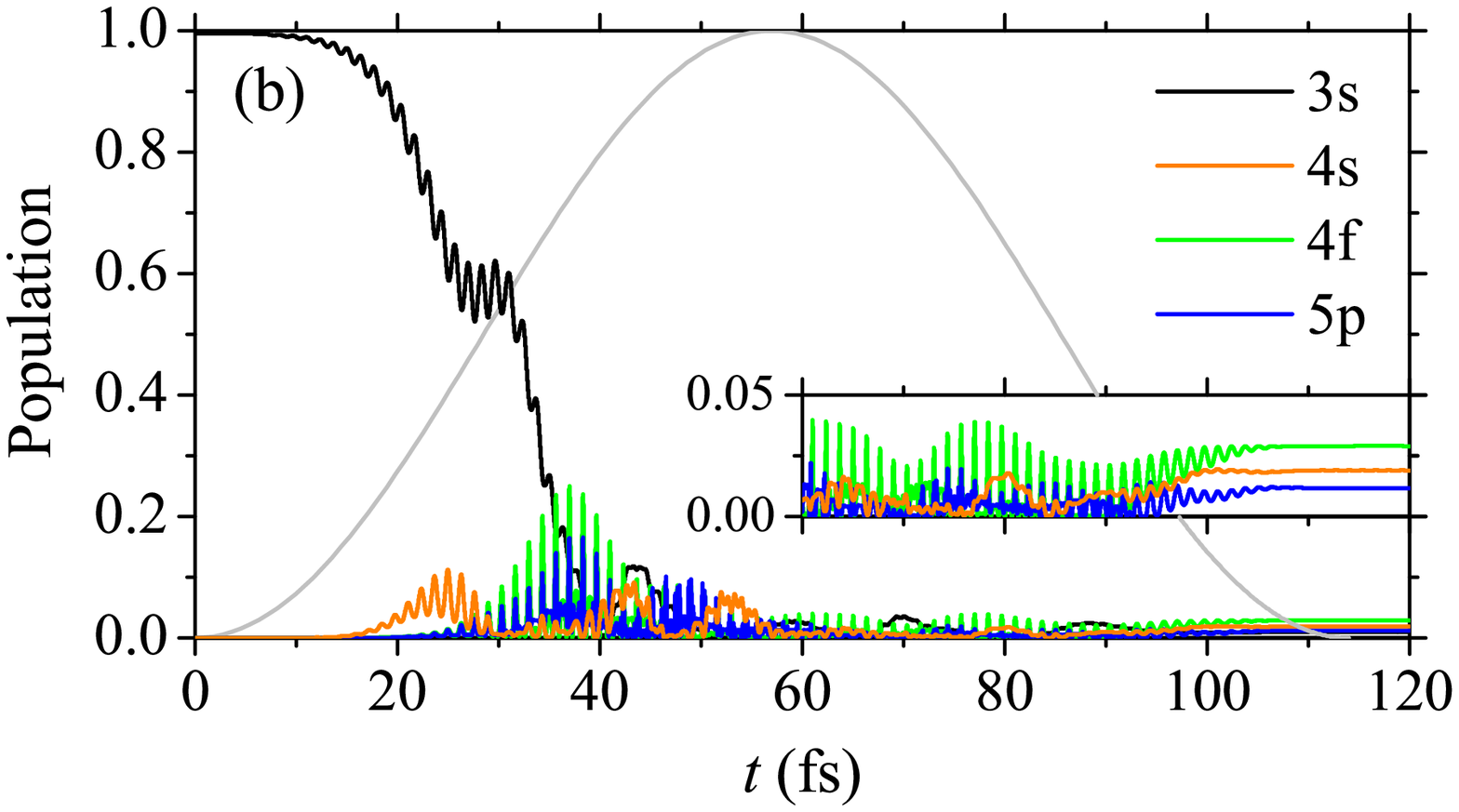}
\\
\includegraphics[width=.45\textwidth]{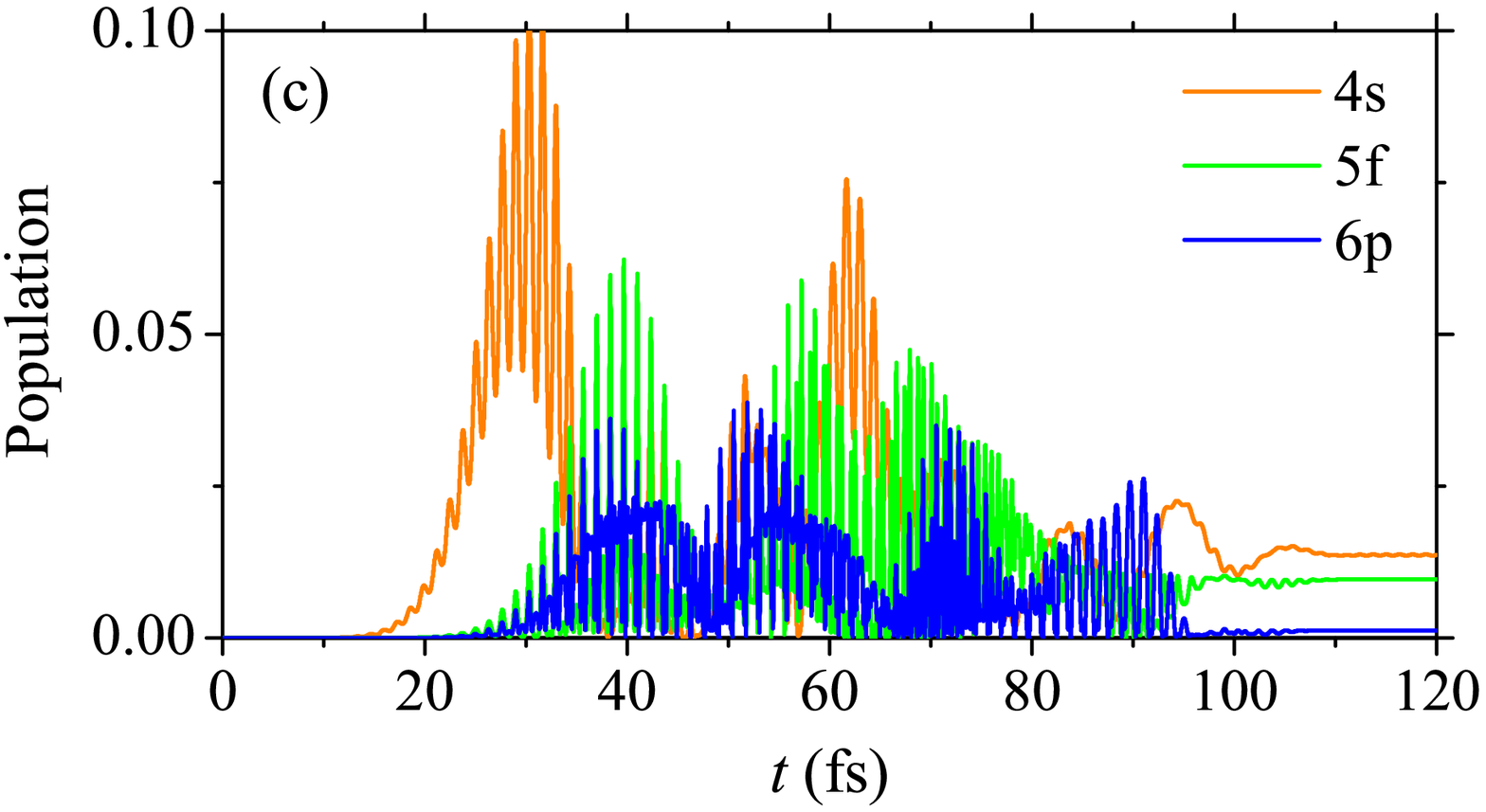}\quad
\includegraphics[width=.45\textwidth]{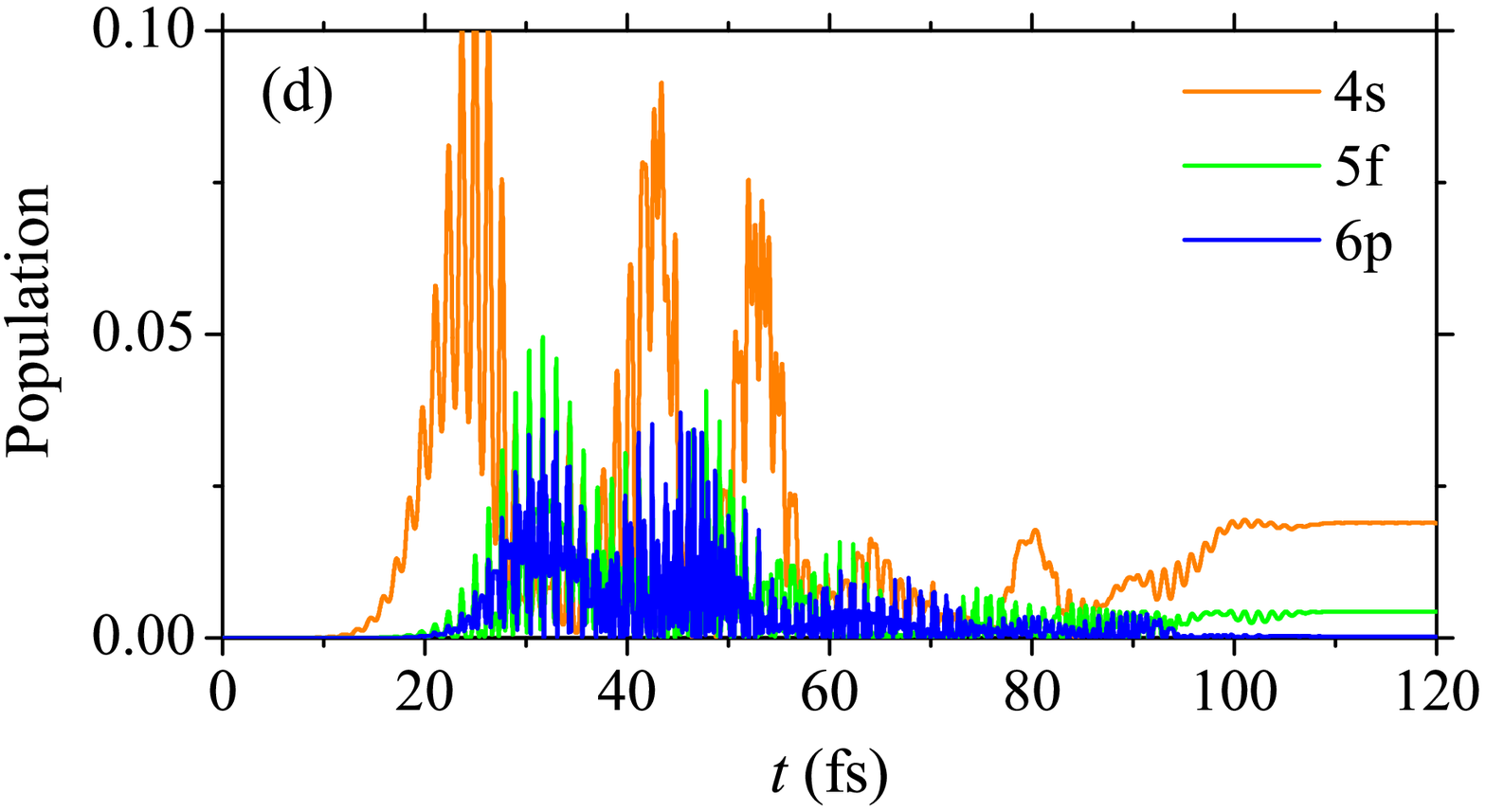}
\\
\includegraphics[width=.45\textwidth]{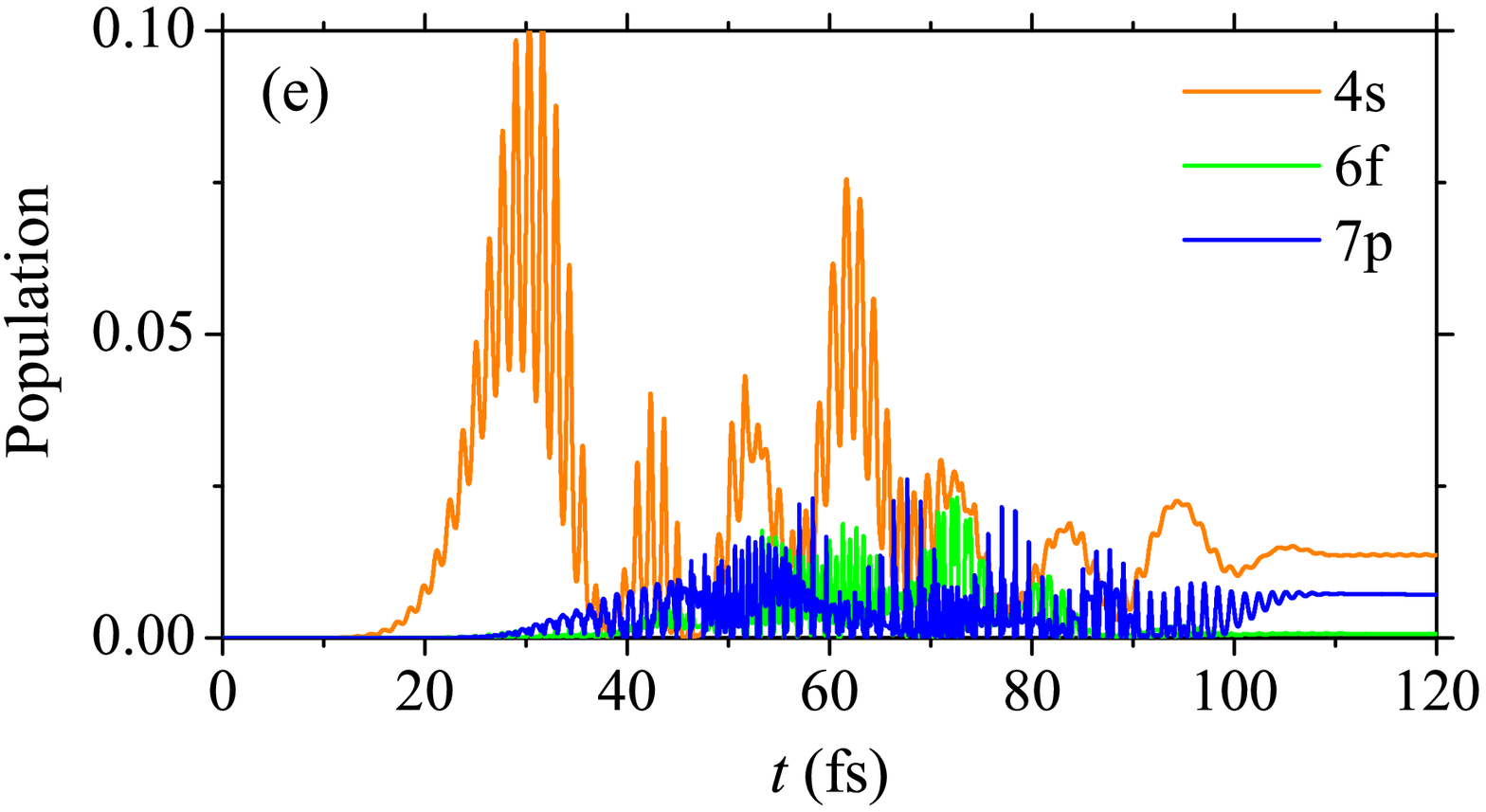}\quad
\includegraphics[width=.45\textwidth]{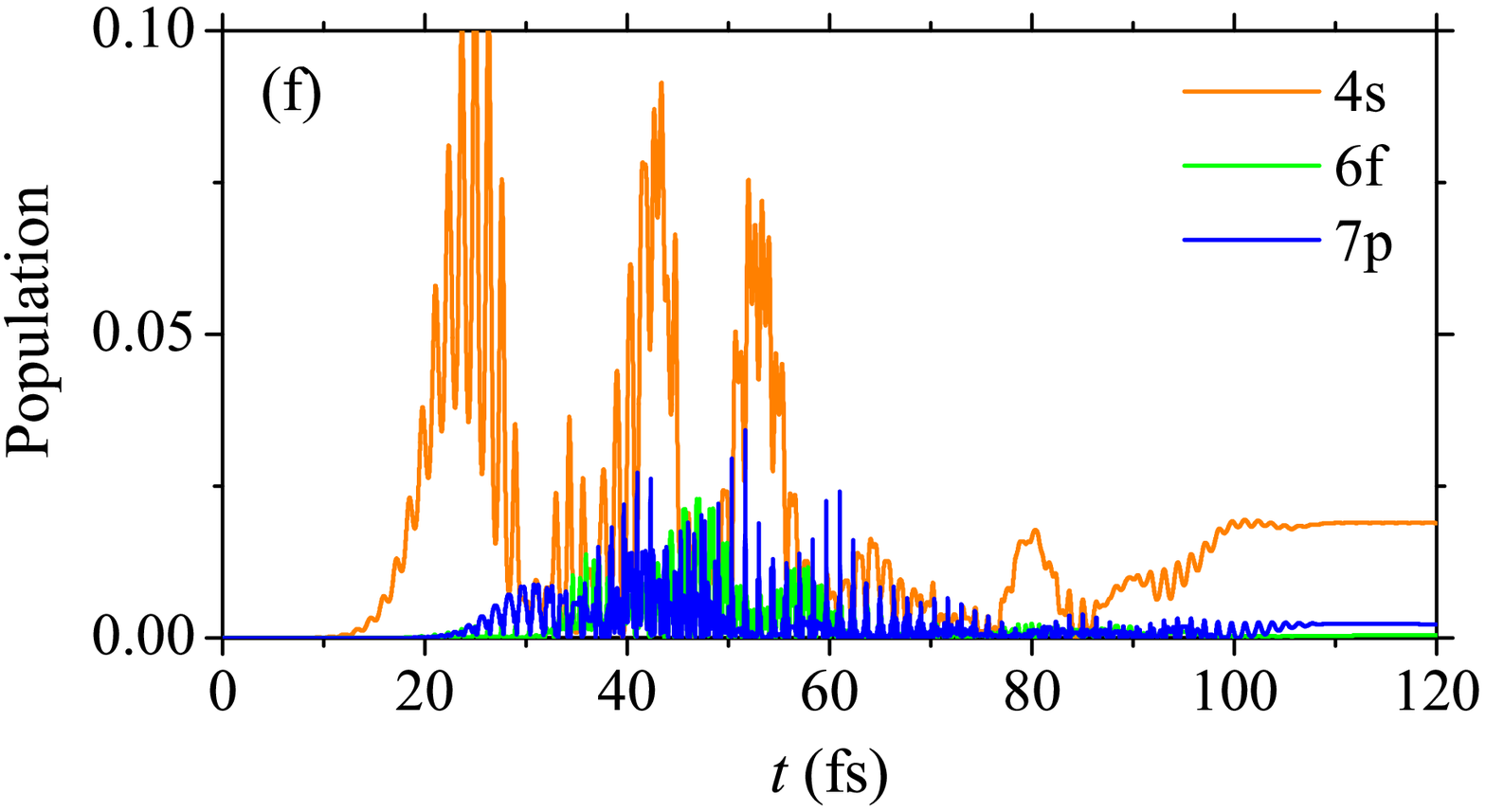}
\end{center}
\caption{Populations of the unperturbed ground (3s) and excited
states 4s, 4f, 5p, 5f, 6p, 6f and 7p of sodium during the 114 fs
laser pulse of 800\,nm wavelength with peak intensities
3.5\,TW/cm$^2$ (left column) and 8.8\,TW/cm$^2$ (right column),
calculated by the wave-packet propagation method. The gray line
represents normalized envelope $F(t)/F_\mathrm{peak}$ of the
pulse.} \label{fig:pop}
\end{figure*}

These findings are verified by calculating the populations of the
unperturbed atomic states $nl$ (i.e. transition probabilities
$|\langle nl|\psi(t)\rangle|^2$) during the pulse, while solving
the TDSE. Fig.~\ref{fig:pop} shows the populations of relevant S,
P and F states during the pulse of 3.5\,TW/cm$^2$ and
8.8\,TW/cm$^2$ peak intensity. Although $I = 3.5\,
\mathrm{TW/cm}^2$ corresponds to the OBI threshold, for this peak
intensity at all phases of the pulse still there is a significant
population of atoms which are not ionized (see the left side of
Fig.~\ref{fig:pop}). The population of states 5p and 7p is here
higher than that of states 4f and 6f, respectively, but the
population of state 5f is higher than the population of state 6p.
Contrarily, at $I = 8.8\,\mathrm{TW/cm}^2$ most atoms after the
first half of the pulse are ionized (see the right side of
Fig.~\ref{fig:pop}). Among the atoms that were not ionized the
population of states 4f and 5f in this case is higher than that of
states 5p and 6p, respectively, while the population of state 7p
is again higher than the population of state 6f. These
observations fully agree with the findings concerning the
photoionization channels.

The populations at different phases of the laser pulse can be well
understood from diagrams (b) and (c) in Fig.~\ref{fig:diagram}.
Diagram (b) predicts and Fig.~\ref{fig:pop}(a,b) confirms that
three-photon transitions from the ground state to states 4f and 5p
are resonant with the radiation of 800\,nm at the field strength
$F \approx 0.01$\,a.u. (see also Table~\ref{table1}), which takes
place in the middle of laser pulse when $I = 3.5\,
\mathrm{TW/cm}^2$ and at $T_p/3$ and $2T_p/3$ for $I = 8.8\,
\mathrm{TW/cm}^2$. States 5f and 6p, on the other hand, shift into
resonance at smaller values of the field strength, which are
reached at the beginning and at the end of pulse (see
Fig.~\ref{fig:pop}(c,d)). Contrarily, the three-photon transitions
from the ground state to states 6f an 7p are not resonant with the
radiation of 800\,nm, but 2+1-photon transition $3\mathrm{s} \to
4\mathrm{s} \to 7\mathrm{p}$ is near resonant in the weak field
limit. Since the dynamic Stark shift for P states grows with the
field strength approximately as the shift for 4s state (see
Fig.~\ref{fig:diagram}(c)), the transition $4\mathrm{s}\to
7\mathrm{p}$ remains near resonant and occurs during the whole
pulse (see Fig.~\ref{fig:pop}(e,f)). This is supported by the fact
that populating the 4s state continues at higher field strengths
due to Rabi oscillations between this and the ground state. In
addition, the transitions $4\mathrm{s}\to 5\mathrm{p}$ and
$4\mathrm{s}\to 6\mathrm{p}$, which also occur due to Rabi
oscillations, partially contribute to the population of 5p and 6p
states.

\section{Summary and conclusions}
\label{sec:conc}

In this paper we studied the photoionization of sodium by
femtosecond laser pulses of 800\,nm wavelength in the range of
field strengths entering over-the-barrier ionization domain and
analyzed possibilities for selective resonantly enhanced
multiphoton ionization through a single channel. Using the
single-active-electron approximation we calculated the
photoelectron momentum distributions by solving numerically the
time dependent Schr\"odinger equation for these pulse parameters.
In order to determine the contribution of different ionization
channels to the total photoelectron yield, a partial wave analysis
of the outgoing wave function in momentum representation is
performed giving the partial probability densities $w_l$ as
functions of the photoelectron energy $\epsilon$. The total
density, which is the sum $\sum_l w_l(\epsilon)$, represents the
photoelectron energy spectrum. The spectra calculated for the
pulses of 800\,nm wavelength, 57\,fs duration (FWHM) and
3.5-8.8\,TW/cm$^2$ peak intensity agree well with the spectra
obtained experimentally by Hart~{\em et al.}~\cite{hart2016}. This
holds for the positions of both REMPI and ATI peaks.

A partial wave analysis of the spectral peaks related to Freeman
resonances has shown that each peak is a superposition of the
contributions of photoelectrons from different ionization
channels. It is found that at the laser peak intensity of
3.5\,TW/cm$^2$ dominant contribution in the main peak (around
0.8\,eV) comes from d-electrons, while at the intensity of
8.8TW/cm$^2$ the electrons of g-symmetry dominate. In this way, by
changing the laser intensity, it is possible to select the main
ionization channel. In the first case this is combined 3+1 and
2+1+1 REMPI via 5p state, while in the second case it is the 3+1
REMPI via 4f state. In contrast to the main peak, the structure of
local peaks around 1\,eV and 1.2\,eV is not sensitive to the laser
intensity. The peak at 1\,eV is related to 3+1 REMPI via the
states 5f and 6p, whereas the dominant ionization channel for the
peak around 1.2\,eV is 2+1+1 REMPI via the near resonant 4s state
and subsequently excited 7p state. These findings are justified by
calculating the populations of excited states during the pulse.
The selectivity might be further improved by choosing more
appropriate pulse parameters which lead to a better resonance
between intermediate states and the laser field.

\medskip

This work was done in the Laboratory for Atomic Collision
Processes, Institute of Physics Belgrade, under Project
No.~OI171020 of the Ministry of Education, Science, and
Technological Development of the Republic of Serbia. Numerical
simulations were run on the PARADOX-IV supercomputing facility at
the Scientific Computing Laboratory, National Center of Excellence
for the Study of Complex Systems, Institute of Physics Belgrade,
supported in part by the Ministry of Education, Science, and
Technological Development of the Republic of Serbia under project
No.~OI171017.

\end{document}